\documentclass[a4paper,11pt]{article}

\usepackage{amsmath}
\usepackage{amsfonts}
\usepackage{amssymb}
\usepackage{amsthm}
\usepackage{hyperref}
\usepackage{xcolor}
\usepackage[numbers,sort&compress]{natbib}

\usepackage[a4paper,left=2.5cm,right=2.5cm,top=2.5cm,bottom=2.5cm, headsep=0.5cm]{geometry}

\newcommand{\vp}{\varphi}
\newcommand{\I}{\mathrm{i}}
\newcommand{\D}{\mathrm{d}}

\renewcommand{\O}{\mathcal{O}}
\newcommand{\<}{\langle}
\renewcommand{\>}{\rangle}
\newcommand{\bs}[1]{\boldsymbol{#1}}
\newcommand{\nn}{\nonumber}
\newcommand{\lla}{\langle \! \langle}
\newcommand{\rra}{\rangle \! \rangle}

\newcommand{\hdelta}{\hat{\delta}}

\begin{document}

\unitlength = 1mm

\begin{center}
\hfill\\
\vspace{1.2cm}
{\LARGE {\bf Comments on scale and conformal invariance\\[10pt]in four dimensions}}

\vspace{1.0cm}
{\large Adam Bzowski$^{a}$, Kostas Skenderis$^{b}$}

\vspace{0.8cm}

{\small
$^a${\it Institute for Theoretical Physics, K.U. Leuven, Belgium.} \\
$^b${\it STAG research centre and Mathematical Sciences, University of Southampton, UK.} \\
}

\vspace{0.5cm}
{\small E-mail: {\tt adam.bzowski@fys.kuleuven.be, k.skenderis@soton.ac.uk} }

\vspace{1.0cm}
\end{center}

\begin{center}
{\bf Abstract}
\end{center}

There has been recent interest in the question of whether four dimensional scale invariant unitary quantum field theories are actually conformally invariant. In this note we present a complete analysis of possible scale anomalies in correlation functions of the trace of the stress-energy tensor in such theories. We find that 2-, 3- and 4-point functions have a non-trivial anomaly while connected higher point functions are non-anomalous. We pay special attention to semi-local contributions to correlators (terms with support on a set containing both coincident and separated points) and show that the anomalies in 3- and 4-point functions can be accounted for by such contributions. We discuss the implications of the our results for the question of scale versus conformal invariance.

\numberwithin{equation}{section}
\pagestyle{empty}
\pagebreak
\setcounter{page}{1}
\pagestyle{plain}
\setcounter{tocdepth}{2}

\tableofcontents

\newpage

\section{Introduction}

It has been a long standing conjecture that every unitary scale invariant quantum field theory (SFT) in four spacetime dimensions is automatically conformally invariant. While in $d=2$ spacetime dimensions scale invariance together with unitarity implies conformal invariance \cite{Zamolodchikov:1986gt,Polchinski:1987dy}\footnote{See however
\cite{Hull:1985rc} for a counterexample where some of assumptions of \cite{Polchinski:1987dy} do not hold.}, the problem has remained open in higher dimensions. Evidence that the conjecture may hold in four dimensions was presented in \cite{Dorigoni:2009ra} while candidate counterexamples \cite{Fortin:2011ks,Fortin:2011sz,Fortin:2012ic} were shown to actually be CFTs in \cite{Luty:2012ww, Fortin:2012hn}. As in the two-dimensional case, there is a connection between RG-flow and
properties of the Wess-Zumino action \cite{Jack:1990eb, Osborn:1991gm} (see also \cite{Jack:2013sha, Baume:2014rla}), the $a$-theorem, \cite{Komargodski:2011vj,Komargodski:2011xv}, and the scale vs conformal invariance problem.
The conjecture is known to hold under some additional assumptions, for example for any scale invariant theory which is obtained by weakly coupled renormalisation group (RG) flows \cite{Luty:2012ww, Baume:2014rla}.
On the other hand, the conjecture does not hold for non-unitary theories
as there are counterexamples \cite{Riva:2005gd}. A less known counterexample is that of topological quantum field theories \cite{Witten:1988ze}. Such theories however do not have local degrees of freedom. There are other counterexamples but all of them are somewhat special: theories without a stress energy tensor \cite{Dorigoni:2009ra}, free $(d-2)$-forms in $d$-dimensions\footnote{Such a form can be dualized to a free scalar $\phi$ with a shift symmetry $\phi \to \phi + const$. The improvement term, $\int R \phi^2$, that would make the theory a CFT  is not compatible with the shift symmetry and thus these theories are scale but not conformally invariant.}, Maxwell theory in $d \neq 4$ \cite{Jackiw:2011vz,ElShowk:2011gz}\footnote{In this case there is a Weyl invariant extension \cite{Deser:1983mm} but the model is not gauge invariant.}, free high-spin theories \cite{Iorio:1996ad}.
Discussions of the conjecture in dimensions other than four can be found, for example, in \cite{Jackiw:2011vz,ElShowk:2011gz}. Early literature on this topic includes \cite{Callan:1970ze,Coleman:1970je} and for
recent reviews (and a more comprehensive list of references) we refer to \cite{Jackiw:2011vz,Nakayama:2013is}.

Recently, two papers \cite{Dymarsky,Luty} argued that the conjecture holds in four spacetime dimensions\footnote{{\bf Note added}: \cite{Luty} was withdrawn after our paper appeared on the arXiv (for the reasons we explain in this paper, see also \cite{Dymarsky:2014zja}).
We will however leave the reference to \cite{Luty} as this provides the context of some of our discussions.}. In \cite{Luty} the structure of the scale anomaly in the 3-point function of the trace of the stress energy tensor was analysed and argued that such anomaly is not consistent with OPEs. We will revisit this argument here and show that the inconsistency disappears after including possible contributions from semi-local terms.  In \cite{Dymarsky} the authors argued
that in SFTs obtained by RG flows an infinite number of matrix element must vanish in a suitable kinematical configuration. The vanishing of these matrix elements is a necessary condition for conformal invariance and the authors argued that it is also a sufficient condition. We attempted to strengthen this argument by combining it with the structure of anomalies. The 4-point function of the trace of stress energy tensor has a non-trivial anomaly which is non-vanishing in the on-shell forward scattering limit. If one were able to show that the 4-point function of the trace of stress energy tensor, including semi-local terms, vanishes in this kinematical limit or that the anomaly cannot be supported by semi-local terms alone then one would conclude that the scale anomaly coefficient must vanish and (as we will argue in detail later) this would imply that the SFT is a CFT. However, the vanishing of the dilaton amplitudes only implies that the  4-point function of the trace of stress energy tensor is semi-local (in the on-shell forward scattering limit) and moreover it turns out that the
anomaly can be supported by semi-local terms alone so one cannot conclude (based on these considerations alone) that the SFT is a CFT.

This paper is organized as follows. In the next section we discuss in more detail the conclusions one can draw for the scale vs conformal problem from the structure of  scale anomalies.  The rest of the paper is devoted to the derivation of the structure of  scale anomalies. In more detail,  after a discussion of the setup in section \ref{sec:setup} we proceed in sections \ref{sec:2pt}, \ref{sec:3pt}, \ref{sec:4pt}, \ref{sec:higher} to analyse the anomaly in 2-, 3-, 4- and higher point functions. In particular, we calculate the most general form of the scale violation in the following 3- and 4-point functions: $\< TTT \>$, $\< TT \O_2 \>$, $\< TT \O_4 \>$, $\< TTTT \>$, where $\O_2$ and $\O_4$ denote operators of dimensions two and four respectively. Since $T = - \partial_\mu V^\mu$, where $V^\mu$ is the virial current (defined in (\ref{e:defV0})), one can compute these correlation functions in two different ways: either directly or by calculating corresponding correlation functions involving the virial current. By comparing the results obtained by the two methods, one can impose strong conditions on the structure of the SFT. We conclude in section \ref{sec:con}.

We relegate many results which are of technical nature and alternative derivations to four appendices. In appendix \ref{sec:large_momentum_limit} we discuss subtleties in the relation between the short-distance/large momentum limit and OPEs in momentum space,
in appendix \ref{sec:p2} we present an alternative derivation of the scale anomaly for the 3- and the 4-point function which does not use the Wess-Zumino action and in appendix \ref{sec:varphi} we compute the anomaly in 3- and 4-point functions using a different parametrisation for the dilaton. This parametrisation has the feature that the contribution of the Wess-Zumino action vanishes and the entire contribution to the scale anomaly is manifestly due to semi-local terms. In appendix \ref{sec:multiple} we discuss a generalisation of our results to the case of the theory containing multiple scalar operators of dimension two and four.

\section{Are unitary scale invariant theories conformal?} \label{sec:intro2}

The standard approach to the problem of enhancing scale invariance to conformal invariance is based on the analysis of improvement terms. In a scale invariant theory the Noether current associated with scale transformations takes form
\begin{equation} \label{e:defV0}
j^\mu = T^\mu_\nu x^\nu + V^\mu,
\end{equation}
where $T_{\mu\nu}$ denotes the stress-energy tensor and $V^\mu$ is called the \emph{virial current}. The conservation of the scale current implies $T = - \partial_\mu V^\mu$. It can be shown \cite{Polchinski:1987dy} that if the virial current is a total derivative, \textit{i.e.}, if
\begin{equation} \label{e:improve}
V^\mu = \partial_\alpha L^{\mu \alpha}
\end{equation}
for some tensor $L^{\mu\alpha}$, then the stress-energy tensor may be redefined to be traceless, hence implying that the theory is conformally invariant.

(Non-anomalous) scale invariance implies that the 2-point function of the trace of the stress-energy tensor is determined up to a constant and is given by
\begin{equation} \label{e:TTnoA}
\< T(\bs{p}) T(\bs{p}') \> = (2 \pi)^4 \delta(\bs{p} + \bs{p}') 2 e_{TT} p^4,
\end{equation}
where $e_{TT}$ is a constant (the factor of 2 is for later convenience).
This correlator however is local and may be removed by a local counterterm (see below). A non-trivial example of a SFT exhibiting this behavior is given by topologically twisted $N=2$ SYM in four dimensions \cite{Witten:1988ze}.
After the topological twist the stress-energy tensor is BRST-trivial explaining the triviality of (\ref{e:TTnoA})
(actually all correlation functions of the stress-energy tensor are trivial). Topological QFTs are special as they do not have local degrees of freedom. In the remaining of this paper we will focus on theories with local degrees of freedom.

In order to obtain a SFT with local excitations and non-zero $T$ (if such a theory exists) the scale transformations must be anomalous. The form of the 2-point function is uniquely fixed by (the now anomalous) scale invariance
and it may be obtained by dimensionally regularising (\ref{e:TTnoA}), $d = 4 - \epsilon$, and taking $e_{TT}$ to be singular in $\epsilon$, $e_{TT} \to e_{TT}/\epsilon$. Expanding in $\epsilon$,
the regulated expression reads
\begin{equation} \label{e:TTreg}
\< T(\bs{p}) T(\bs{p}') \>_{\text{reg}} = (2 \pi)^4 \delta(\bs{p} + \bs{p}') \left[ \frac{2 e_{TT} p^4}{\epsilon} - e_{TT} p^4 \log p^2 + O(\epsilon) \right],
\end{equation}
The 2-point function now requires renormalisation and the divergence may be removed by the counterterm
\begin{equation} \label{e:TTct}
S_{ct} = \left( \frac{e_{TT}}{\epsilon} + e_{TT}^{(0)} \right) \int \D^{4 - \epsilon} \bs{x} \tfrac{1}{36} R^2 \mu^{-\epsilon}.
\end{equation}
The value of the divergent term is directly related to the normalisation constant of the 2-point function while the value of the finite piece may be adjusted as will.
This leads to the renormalised correlation function
\begin{equation} \label{e:TT0}
\< T(\bs{p}) T(\bs{p}') \> = (2 \pi)^4 \delta(\bs{p} + \bs{p}') \left[ - e_{TT} p^4 \log \frac{p^2}{\mu^2} + e_{TT}^{\text{loc}}(\mu) p^4 \right],
\end{equation}
where $e_{TT}^{\text{loc}}(\mu)$ is a scheme dependent constant.

The scale (or dilatation) symmetry may be gauged and the resulting theory becomes classically Weyl invariant. The counterterm \eqref{e:TTct} is not Weyl invariant and requires the addition of appropriate terms. The Weyl invariant form of the counterterm is
\begin{equation}
S_{ct} = \left( \frac{e_{TT}}{\epsilon} + e_{TT}^{(0)} \right) \int \D^{4 - \epsilon} \bs{x} \left( \tfrac{1}{6} R + \nabla_{\alpha} C^{\alpha} - C_{\alpha} C^{\alpha} \right)^2 \mu^{-\epsilon},
\end{equation}
where $C_\mu$ denotes the source for the virial current $V^\mu$ and under Weyl transformations one has $\delta_\sigma C_{\mu} = \partial_\mu \sigma$. At the quantum level the Weyl symmetry is anomalous.

The anomaly can be represented in terms of the Wess-Zumino action, \textit{i.e.}, one can divide the generating functional of connected graphs $W$ into a Weyl invariant part $W_{WI}$ and an anomalous part $W_A$,
$W = W_{WI} + W_{A}$,
\begin{equation}
W_A[e^{2 \sigma} g_{\mu \nu}, C_\mu + \partial_\mu \sigma, \dots] = W_A[g_{\mu \nu}, C_\mu, \dots] + S_{WZ}[g_{\mu \nu}, C_\mu, \dots; \sigma]
\end{equation}
where the dots indicate sources for operators other than $T_{\mu \nu}$ and $V^\mu$.
The most general parity-even form of the Wess-Zumino action involving the metric and the gauge field $C_{\mu}$ reads \cite{Luty:2012ww,Luty},
\begin{align} \label{e:WZ0}
S_{WZ}[g_{\mu \nu}, C_\mu; \sigma] & = \int \D^4 \bs{x} \sqrt{g} \left\{ - a \left[ \sigma E_4 + 4 \left( R^{\mu\nu} - \tfrac{1}{2} g^{\mu\nu} R \right) \partial_\mu \sigma \partial_\nu \sigma - 4 (\partial\sigma)^2 \Box \sigma + 2 (\partial \sigma)^4 \right] \right. \nn\\
& \quad\qquad \left. + \: c \sigma W^2 - e \sigma \Sigma^2 + f \sigma C_{\mu\nu} C^{\mu\nu} \right\},
\end{align}
where $\sigma$ is the dilaton, $E_4$ denotes the Euler density, $W^2$ is the square of the Weyl tensor and
\begin{align}
\Sigma & = \tfrac{1}{6} R + \nabla_\mu C^\mu - C_\mu C^\mu, \\
C_{\mu\nu} & = \partial_\mu C_\nu - \partial_\nu C_\mu.
\end{align}
The coefficients $a$ and $c$ are the standard conformal anomaly coefficients of a CFT. The coefficient $f$ is also a standard CFT anomaly coefficient due to conserved currents. What is new in SFTs that are not CFTs is the $e$ anomaly.

We now want to relate the coefficients in the Wess-Zumino action with coefficients in correlation functions.
On one hand, the Wess-Zumino action may be related to the anomaly in the Weyl Ward identity. With sources for operators other than the stress-energy tensor and the virial current turned off, the identity reads
\begin{align}
\delta_\sigma S_{WZ} = \delta_\sigma W & = \int \D^4 \bs{x} \sqrt{g} \sigma \left( - 2 g^{\mu\nu} \frac{\delta}{\delta g^{\mu\nu}} - \nabla_\mu \frac{\delta}{\delta C_\mu} \right) W \nn\\
& = \int \D^4 \bs{x} \sqrt{g} \sigma \< T + \nabla_\mu V^\mu \>,
\end{align}
where $\delta_\sigma S_{WZ}$ denotes the terms in $S_{WZ}$ that are linear in $\sigma$.
On the other hand, one finds
\begin{align} \label{e:dTT}
& \delta_\sigma \< T(\bs{x}_1) T(\bs{x}_2) \> = 8 \sigma \< T(\bs{x}_1) T(\bs{x}_2) \> + \nn\\
& \qquad \left. + \:  \frac{-2}{\sqrt{g(\bs{x}_1)}} g^{\mu\nu}(\bs{x}_1) \frac{\delta}{\delta g^{\mu \nu}(\bs{x}_1)} \left( \frac{-2}{\sqrt{g(\bs{x}_2)}} g^{\rho\sigma}(\bs{x}_2) \frac{\delta}{\delta g^{\rho\sigma}(\bs{x}_2)} \delta_\sigma S_{WZ} \right) \right|_{g_{\mu\nu} = \delta_{\mu\nu}}.
\end{align}
The first term captures the classical scaling of the 2-point function (in a flat background $\< T(\bs{x}_1) T(\bs{x}_2) \> \sim |\bs{x}_1 - \bs{x}_2|^{-8}$) while the second term represents the scale violation of the 2-point function. By evaluating this expression in momentum space one finds that the $e$-anomaly is equal to the normalisation constant \eqref{e:TT0} of the 2-point function $\< T T \>$ in the SFT, $e = e_{TT}$.

While in conformal theories $e = 0$, in SFTs $e$ may be \textit{a priori} non-vanishing. The converse also holds: if $e = e_{TT} = 0$, then in unitary theories $T = 0$ and the scale invariant theory becomes fully conformal. Thus, a sufficient and necessary condition for a SFT to a CFT is that $e_{TT}=0$\footnote{If the SFT has a dimension two operator $\O_2$ then the condition is, $e_{TT}=e_{2T}^2/e_{22}$, where $e_{2T}$ and $e_{22}$ are normalisation in the 2-point functions of $T$ and $\O_2$, see (\ref{e:2T}) and (\ref{e:22}). When this condition holds one may improve $T$ such that the new $T$ vanishes, see the discussion in section \ref{sec:improv}.}.

In this paper we analyse properties of the SFTs in momentum space. For correlation functions of scalar operators $\O_1, \ldots, \O_n$ of dimensions $\Delta_1, \ldots, \Delta_n$ scale invariance implies
\begin{equation}
\< \O_1(e^\sigma \bs{p}_1) \ldots \O_n(e^\sigma \bs{p}_n) \> = e^{ \left[ \sum_{j=1}^n \Delta_j - n d \right] \sigma} \< \O_1(\bs{p}_1) \ldots \O_n(\bs{p}_n) \> + \mathcal{A}_n(\sigma),
\end{equation}
as we will discuss in the following section. When expanded, the leading term in $\sigma$ consists of two parts: a classical part and an anomalous part, as in \eqref{e:dTT}. Since we are interested in the anomalous term $\mathcal{A}_n(\sigma)$, we may remove the classical piece by defining infinitesimal scale transformation
\begin{equation}
\hdelta_\sigma  \< \O_1(\bs{p}_1) \ldots \O_n(\bs{p}_n) \> = \left. \frac{\D}{\D \sigma} \left( e^{ - \left[ \sum_{j=1}^n \Delta_j - n d \right] \sigma} \< \O_1(e^\sigma \bs{p}_1) \ldots \O_n(e^\sigma \bs{p}_n) \> \right) \right|_{\sigma=0},
\end{equation}
which picks up the leading term in $\sigma$ from the anomaly $\mathcal{A}_n(\sigma)$. If no anomalies are present, $\hdelta_\sigma \< \O_1(\bs{p}_1) \ldots \O_n(\bs{p}_n) \> = 0$.

Using the relation between the Wess-Zumino action and \eqref{e:dTT} it was argued in \cite{Luty} that $e_{TT} = 0$ in any SFT, due to the consistency between the Wess-Zumino action and the OPE of the stress-energy tensor. As we will show in section \ref{sec:TTT}, the scale violation of the 3-point function of the trace of the stress-energy tensor is uniquely determined and reads
\begin{equation} \label{e:TTT0}
\hdelta_\sigma \< T(\bs{p}_1) T(\bs{p}_2) T(\bs{p}_3) \> = (2 \pi)^4 \delta (\bs{p}_1 + \bs{p}_2 + \bs{p}_3) \times 2 \sigma e_{TT} J^2,
\end{equation}
where
\begin{equation}
J^2 = -p_1^4 - p_2^4 - p_3^4 + 2 p_1^2 p_2^2 + 2 p_1^2 p_3^2 + 2 p_2^2 p_3^2.
\end{equation}
In deriving \eqref{e:TTT0} we used the fact (derived in section \ref{sec:imp}) that one can always add an appropriate improvement term such that
 all off-diagonal 2-point functions of the trace of the stress-energy tensor and scalar operators $\O_2, \O_4$ of dimensions two and four vanish,
\begin{equation} \label{e:offdiag0}
\< T(\bs{p}) \O_2(\bs{p}') \> = \< T(\bs{p}) \O_4(\bs{p}') \> = \< \O_2(\bs{p}) \O_4(\bs{p}') \> = 0.
\end{equation}

In \cite{Luty} the following OPE argument was used to argue that such a scale violation in the 3-point function is not possible in any SFT. The argument is based on the observation that in position space the OPE implies that for $\bs{x}_1 \rightarrow \bs{x}_2$,
\begin{equation} \label{e:OPE}
\< T(\bs{x}_1) T(\bs{x}_2) T(\bs{x}_3) \> \sim \frac{1}{|\bs{x}_1 - \bs{x}_2|^{8 - \Delta_\O}} \< \O(\bs{x}_2) T(\bs{x}_3) \> + \frac{x_1^\mu - x_2^\mu}{|\bs{x}_1 - \bs{x}_2|^{9 - \Delta_K}} \<  K_\mu(\bs{x}_2) T(\bs{x}_3)\> + \ldots
\end{equation}
where $\O$ and $K_\mu$ are the scalar and vector operators of the lowest dimension contributing to the OPE. Then the Fourier transform of this expression is compared to the large momentum limit $p_1 \cong p_2 \gg p_3$ of \eqref{e:TTT0}. It is argued that the momentum dependence obtained via the OPE is such that the scale violation of $\< TTT \>$
cannot match (\ref{e:TTT0}) and thus $e_{TT}$ must be equal to zero. Hence the theory is conformal.

The critical flaw in this argument is the assumption that the large momentum limit $q = p_1 \cong p_2 \gg p_3 = p$ follows directly from the Fourier transform of the leading $\bs{x}_1 \rightarrow \bs{x}_2$ behaviour in \eqref{e:OPE}. In appendix \ref{sec:large_momentum_limit} we argue that the correct large momentum expansion reads
\begin{equation} \label{e:op0}
\< \O_1(\bs{q}) \O_2(-\bs{q}+\bs{p}) \O_3(-\bs{p}) \> \propto \left\{ \begin{array}{ll}
q^{\Delta_1 + \Delta_2 - \Delta_3 - d} p^{2 \Delta_3 - d} \left( 1 + o \left(p/q \right) \right) & \text{if } \Delta_3 < \frac{d}{2} \\
q^{\Delta_1 + \Delta_2 + \Delta_3 - 2 d} \left( 1 + o \left(p/q \right) \right) & \text{if } \Delta_3 > \frac{d}{2} \end{array} \right.
\end{equation}
up to proportionality factors. In special cases such as $2 \Delta_3 = d$, logarithms can appear as well. We will not list all  possibilities since we are interested in explaining why the two forms appear in the generic case.

The behaviour presented in the first line of \eqref{e:op0} will be called \emph{a naive OPE} behaviour, since it follows directly from the Fourier transform of the appropriate OPE term. Indeed, if the OPE reads
\begin{equation} \label{e:ope1}
\O_1(\bs{x}_1) \O_2(\bs{x}_2) \sim \frac{C_{123}}{|\bs{x}_1 - \bs{x}_2|^{\Delta_1 + \Delta_2 - \Delta_3}} \O_3(\bs{x}_2) + \ldots
\end{equation}
then it would be natural to expect that the leading large-momentum behaviour follows from the Fourier transform of the leading $\bs{x}_1 \rightarrow \bs{x}_2$ behaviour of the 3-point function
\begin{equation}
\< \O_1(\bs{x}_1) \O_2(\bs{x}_2) \O_3(\bs{x}_3) \> \sim \frac{C_{123}}{|\bs{x}_1 - \bs{x}_2|^{\Delta_1 + \Delta_2 - \Delta_3}} \< \O_3(\bs{x}_2) \O_3(\bs{x}_3) \>.
\end{equation}
The Fourier transform of this expression leads to the first line of \eqref{e:op0}, since
\begin{equation} \label{e:2ptFourier}
\int \D^d \bs{x} \: e^{-\I \bs{p} \cdot \bs{x}} \frac{1}{x^{2 \Delta}} = \frac{\pi^{d/2} 2^{d - 2 \Delta} \Gamma \left( \frac{d - 2 \Delta}{2} \right)}{\Gamma ( \Delta )} p^{2 \Delta - d}.
\end{equation}

This reasoning however is incorrect in general. While there is an interplay between the large momentum limit $p_1, p_2 \gg p_3$ and the coincident limit $\bs{x}_1 \rightarrow \bs{x}_2$, it is not as straightforward as suggested by the naive OPE argument. Nevertheless, observe that the term $q^{\Delta_1 + \Delta_2 + \Delta_3 - 2 d}$ in the second line of \eqref{e:op0} is semi-local, \textit{i.e.}, up to a constant it is a Fourier transform of the expression
\begin{equation} \label{e:locterm1}
\delta(\bs{x}_2 - \bs{x}_3) \frac{1}{| \bs{x}_1 - \bs{x}_3|^{\Delta_1 + \Delta_2 + \Delta_3 - d}} \quad \stackrel{\mathcal{F}}{\longmapsto} \quad p_1^{\Delta_1 + \Delta_2 + \Delta_3 - 2 d}
\end{equation}
and hence it is a Fourier transform of the distribution supported on the set of coincident points. As was pointed out in \cite{Maldacena:2011nz} (and we discuss in detail in appendix \ref{sec:3k}), the first line of \eqref{e:op0} represents the first \emph{non-local} term in all cases, \textit{i.e.}, the term that in position space is not supported on a set of coincident points.

Returning to the anomaly, notice that all terms in \eqref{e:TTT0} can originate from Fourier transforms of semi-local expressions, for example
\begin{align}
& \delta(\bs{x}_2 - \bs{x}_3) \frac{1}{| \bs{x}_1 - \bs{x}_3|^{8}} \quad \stackrel{\mathcal{F}}{\longmapsto} \quad p_1^4 \log p_1^4, \nn\\
& \Box_{\bs{x}_2} \delta(\bs{x}_2 - \bs{x}_3) \frac{1}{| \bs{x}_1 - \bs{x}_3|^{6}} \quad \stackrel{\mathcal{F}}{\longmapsto} \quad p_1^2 p_2^2 \log p_1^4.
\end{align}
Including such semi-local terms one finds that the Weyl variation of $\<TTT\>$ can indeed match the anomaly obtained from the WZ action and one cannot conclude that $e_{TT}=0$.

A different approach to the problem of enhancing scale invariance to conformal invariance was undertaken in \cite{Dymarsky}. The authors analysed dilaton amplitudes defined as
\begin{equation} \label{e:An}
A_n = \frac{\delta^n W}{\delta \varphi(\bs{x}_1) \ldots \delta \varphi(\bs{x}_n)},
\end{equation}
where $\varphi$ is the scale mode of the metric, $g_{\mu\nu} = (1 + \varphi)^2 \delta_{\mu\nu}$. The dilaton $\varphi$ is a source for the trace of the stress-energy tensor in the sense that it couples to $T$, $S_{\text{int}} = - \int \varphi T + O(\varphi^2)$. Then they argued that the imaginary part of these amplitudes must vanish in an on-shell $p_j^2 \rightarrow 0$ and forward kinematics limit. Using the optical theorem they then concluded that the entire amplitudes $A_n$ must vanish in this kinematical limit (assuming this limit exists, see \cite{Luty:2012ww} and \cite{Baume:2014rla} for a discussion of this point for $A_4$ \footnote{These papers used OPEs in order to control the behavior of the amplitude in momentum space. This raises the question of whether the subtleties we uncover in the relation between OPEs and limits in momentum space would affect their argument. While answering this question in full requires additional study we note that the potentially dangerous contributions come from operators of dimension $\Delta \leq d/2=2$ and for those the naive OPE behavior provides the correct large momentum limit in 3-point functions, see (\ref{e:op0}) (as noted above $\Delta=2$ is special).}).
This then suggests that the interaction terms between the dilatons can be removed by a field redefinition and this would be possible if there exists a local operator $\O_2$ such that $T^\mu_\mu = \Box \O_2$, concluding that the SFT is a CFT.

While this argument is very suggestive it would be preferable to have a more clear-cut proof. As mentioned earlier, a necessary and sufficient condition for a SFT to be a CFT is that the anomaly coefficient $e_{TT}$ vanishes, so one may wonder whether the vanishing of the dilaton amplitudes can be used to show that $e_{TT}=0$. The imaginary part of $A_n$ in the SFT should come from logarithmic terms. Thus if there is a non-trivial scale anomaly which is proportional to $e_{TT}$ one may hope that the vanishing of the imaginary part of $A_n$ would imply $e_{TT}=0$.

We show in section \ref{sec:higher} that there is no anomaly for connected 5- and higher point functions of $T$. Furthermore, 3-point functions are trivial on-shell. Thus, we are left to discuss 4-point functions. It turns out the anomaly for 4-point function is non-trivial and is given by
\begin{align} \label{e:TTTT0}
& \hdelta_\sigma \lla T(\bs{p}_1) T(\bs{p}_2) T(\bs{p}_3) T(\bs{p}_4) \rra = - 8 \sigma \left( e_{TT} + \tfrac{1}{4} c_2^2 e_{22} \right) \times \nn\\
& \qquad\qquad \times \left[ (\bs{p}_1 \cdot \bs{p}_2) (\bs{p}_3 \cdot \bs{p}_4) + (\bs{p}_1 \cdot \bs{p}_3) (\bs{p}_2 \cdot \bs{p}_4) + (\bs{p}_1 \cdot \bs{p}_4) (\bs{p}_2 \cdot \bs{p}_3) \right].
\end{align}
The constant $e_{22}$ is the normalisation of the 2-point function of an operator $\O_2$ of dimension two and $c_2$ is a constant appearing in coupling of $\O_2$ with the dilaton and $C_\mu$, see (\ref{e:22}) and \eqref{e:Sint2}.
In particular, $e_{22} \geq 0$ in any reflection positive theory and hence for the scale violation in the 4-point function to vanish, one necessarily needs $e_{TT} = 0$. Here again we used the fact that the off-diagonal 2-point functions in \eqref{e:offdiag0} may be set to zero by adding improvement terms. Note that this anomaly is non-vanishing in the on-shell forward kinematics limit.

We now explain that despite the fact that the dilaton amplitude $A_4$ vanishes, one cannot conclude that $e_{TT}=0$. Recall that the trace of the stress-energy tensor is the operator defined as
\begin{equation}
T = \frac{2}{\sqrt{g}} g^{\mu\nu} \frac{\delta S}{\delta g^{\mu\nu}},
\end{equation}
and hence it is a functional of the metric and other sources as well. In particular,
\begin{equation} \label{e:funder}
\< T(\bs{x}_1) \ldots T(\bs{x}_n) \> = (-1)^n \frac{\delta^n W}{\delta \varphi(\bs{x}_1) \ldots \delta \varphi(\bs{x}_n)} + \text{semi-local terms},
\end{equation}
and the semi-local terms cannot be disregarded as explained in section \ref{sec:TTTT}. It follows that the vanishing of the dilaton amplitude
in this kinematical limit only implies that the correlators are purely semi-local (in this kinematical configuration). Moreover, the anomaly (\ref{e:TTTT0}) can be completely accounted by semi-local terms, as it is clear from the computation in appendix \ref{sec:varphi} where all contributions come from semi-local terms.

While our analysis does not invalidate the reasoning of \cite{Dymarsky}, we did not manage to provide additional support for it.
Of course, if one accepts that $T=\Box \O_2$ then it suffices to look at the anomaly of the 3-point function to conclude $e_{TT}=0$.
This is so because for the anomalies to match the 3-point function of $\O_2$ would need to have a non-local scale anomaly.

\section{Set-up} \label{sec:setup}

\subsection{Notation and kinematics}

In the paper we work in Euclidean signature. We use bold letters to denote vectors, \textit{e.g.}, $\bs{x}$, $\bs{p}$ and we define $x = | \bs{x} |$, $p = |\bs{p}|$ and so on. Due to the momentum conservation, any correlation function in momentum space carries a delta function. We use the double bracket notation $\lla - \rra$ to denote the omission of this delta function, \textit{i.e.},
\begin{equation}
\< \O_1(\bs{p}_1) \ldots \O_n(\bs{p}_n) \> = (2 \pi)^d \delta \left( \sum_{j=1}^n \bs{p}_j \right) \lla \O_1(\bs{p}_1) \ldots \O_n(\bs{p}_n) \rra.
\end{equation}
Despite the fact that the $n$-point function has $n$ momenta listed as its arguments, only $n-1$ momenta is independent, since $\sum_{j=1}^n \bs{p}_j = 0$.

Due to Lorentz invariance, every 3-point function can be regarded as a function of magnitudes of the three momenta $p_j = |\bs{p}_j|$, $j=1,2,3$.
We will often encounter the following combination of momenta,
\begin{equation} \label{e:J2}
J^2 = -p_1^4 - p_2^4 - p_3^4 + 2 p_1^2 p_2^2 + 2 p_1^2 p_3^2 + 2 p_2^2 p_3^2 = 4 \cdot \text{Gram}(\bs{p}_1, \bs{p}_2),
\end{equation}
where $\text{Gram}$ is the Gram determinant. For physical momentum configurations obeying the triangle inequalities we have $J^2 \geq 0$, with $J^2 = 0$ holding if and only if the three momenta are collinear.

In the case of 4-point functions in $d \geq 4$ spacetime dimensions, the correlators depend on six scalars, which may be taken to be the scalar products $p_{ij} = \bs{p}_i \cdot \bs{p}_j$, $i, j = 1,2,3,4$, and $i \neq j$. While such a parametrisation is the most symmetric one, one can consider other variables, for example four squares of momenta, $p_j^2$, $j=1,2,3,4$ and two Mandelstam variables, say $s = (\bs{p}_1 + \bs{p}_2)^2$ and $t = (\bs{p}_1 + \bs{p}_3)^2$. Such variables turn out to be useful, since the calculations simplify significantly in the forward scattering limit
\begin{equation} \label{e:fwd}
t \rightarrow 0, \qquad p_j^2 \rightarrow 0, \ j=1,2,3,4\, .
\end{equation}

\subsection{Generating functional}

In this paper we consider a reflection positive SFT. The set of operators that can mix with the trace of the stress-energy tensor
are the virial current $V^{\mu}$ and scalar operators $\O_2$ and $\O_4$ of dimensions two and four, respectively. By the unitarity bounds in SFT \cite{Grinstein:2008qk,Luty}, the dimensions $\Delta_s$ of operators of spin $s$ are bounded from below by
\begin{equation} \label{e:bounds}
\Delta_s \geq \left\{ \begin{array}{ll}
1 & \text{ for } s = 0, \\
2 & \text{ for } s = 1, \\
3 & \text{ for } s = 2 \end{array} \right.
\end{equation}
Furthermore, a scalar operator $\Phi$ of dimension one is necessarily a fundamental scalar field, since its 2-point function satisfies $\Box_{\bs{x}} \< \Phi(\bs{x}) \Phi(\bs{y}) \> = 0$. Since we can always decouple a free theory (and moreover a free theory can be improved to be conformal) we assume that $\Delta_0 > 1$ for all scalar operators. In general the dilatation operator may not be diagonalizable due to renormalization effects.

The SFT may be coupled to sources: the metric $g^{\mu\nu}$, the source $C_\mu$ for the virial current, and scalar sources $\phi_2$ and $\phi_0$ for the operators $\O_2$ and $\O_4$, respectively, and this has the effect of gauging the scaling symmetry. In particular, under Weyl transformations
\begin{equation} \label{weyl}
\delta_\sigma g_{\mu\nu} = 2 \sigma g_{\mu\nu},  \qquad \delta_\sigma \phi_{d - \Delta} = - (d - \Delta) \sigma \phi_{d - \Delta}, \qquad \delta_\sigma C_\mu = \partial_\mu \sigma.
\end{equation}
The subscript on the scalar sources denotes their scaling dimensions.
The source $C_\mu$ couples to the dynamical objects via covariant derivatives,
\begin{equation}
\partial_\mu \O \mapsto D_\mu \O = \left( \partial_\mu + \Delta_\O C_\mu \right) \O,
\end{equation}
where $\Delta_{\O}$ is a scaling dimension of the operator $\O$. The transformation property of the gauge field $C_{\mu}$ implies that
$D_\mu$ is a covariant derivative for scale transformations and thus, for example,
$D_\mu \O$ transforms as a field with weight $\Delta_{\O}$ under Weyl transformations.

In this paper we assume that the dilatation operator is diagonalizable and the Weyl transformation rules are given by
(\ref{weyl}). As discussed in \cite{Polchinski:1987dy} in general dilatations may not be diagonalizable and the
most general local transformations that are consistent with the Wess-Zumino consistency condition may contain
additional terms \cite{Osborn:1991gm, Baume:2014rla}. It would be interesting to extend our analysis to the general case.

In \cite{Luty:2012ww,Luty} the most general form of scaling anomalies in the stress-energy tensor and the virial currents in a SFT was obtained. If $W$ denotes the generating functional of connected correlators for the SFT with gauged scaling symmetry, then its Weyl transformation $\delta_\sigma W$ can be expressed as a variation of the local Wess-Zumino action $\delta_\sigma S_{WZ}$. In this paper we need to include operators of dimension two and four and such operators contribute to scaling anomalies \cite{Osborn:1991gm,Petkou:1999fv}.

We are interesting in computing the anomaly of the following correlation functions:
\begin{itemize}
\item 2-point functions of the trace of the stress-energy tensor $T$, the longitudinal part of the virial current $V^\mu$ and scalar operators of dimensions two and four.
\item 3-point functions of $T$ and the longitudinal part of $V^\mu$ with up to a single insertion of the scalar operators.
\item 4-point functions of $T$ and the longitudinal part of $V^\mu$ only.
\end{itemize}
An analysis of 2-point functions analogous to the one in section \ref{sec:intro2} implies that there are scale anomalies in all possible 2-point functions. This implies that we should include all possible dimension 4 terms that are quadratic in the sources with arbitrary coefficients. The values of these coefficients are related to the normalisation constants of 2-point functions, as we will discuss in section \ref{sec:2pt}. Turning to cubic coupling now, we note that we are only interested in 3-point functions with a single insertion of a scalar operator, thus the relevant cubic couplings should be at most linear in $\phi_0$ or $\phi_2$. By dimensional analysis the relevant terms in the Wess-Zumino action read
\begin{align} \label{e:preWZ}
\delta_\sigma S_{WZ} & = \int \D^4 \bs{x} \sqrt{g} \sigma \left[ ( - e_{TT} + e_{4TT} \phi_0 ) R^2 - e_{E44} \phi_0 E_4 \right.\nn\\
& \left.\qquad - \: e_{22} \phi_2^2 - e_{44} (\Box \phi_0)^2 + 2 e_{2T} \phi_2 R + 2 e_{24} \phi_2 \Box \phi_0 + 2 e_{4T} R \Box \phi_0 + \ldots \right].
\end{align}
Since we are interested in the correlation functions of the trace of the stress-energy tensor rather than the entire stress-energy tensor, the square of the Weyl tensor may be omitted. Furthermore, since the Euler density is a topological term, it does not contribute to the 2-point function of $T$.

By introducing the gauge field $C_\mu$, one can gauge the Weyl transformations and extend the action \eqref{e:preWZ} to be fully Weyl-invariant. To this end let us define,
\begin{align}
\hat{R} = 6 \Sigma & = R + 6 \nabla_\alpha C^\alpha - 6 C_\alpha C^\alpha, \\
\hat{R}_{\mu\nu} & = R_{\mu\nu} + 2 \nabla_{(\mu} C_{\nu)} + g_{\mu\nu} \nabla_\alpha C^\alpha + 2 C_{\mu} C_{\nu} - 2 g_{\mu\nu} C_\alpha C^\alpha, \\
\hat{\Box} & = \Box - 2 C^{\alpha} \nabla_{\alpha}
\end{align}
and note that these objects transform homogeneously under Weyl transformations,
\begin{equation}
\delta_\sigma \hat{R} = - 2 \sigma \hat{R}, \qquad \delta_\sigma \hat{R}_{\mu\nu} = 0, \qquad \delta_\sigma (\hat{\Box} f) = - 2 \sigma \hat{\Box} f + \hat{\Box}(\delta_\sigma f).
\end{equation}
It follows that one can produce gauge invariant quantities by replacing un-hatted by hatted quantities in \eqref{e:preWZ}. In particular, the gauged Euler density reads
\begin{equation}
\hat{E}_4 = W^2 - 2 \hat{R}_{\mu\nu}^2 + \tfrac{2}{3} \hat{R}^2,
\end{equation}
where $W^2$ is the usual square of the Weyl tensor, which transforms homogeneously under Weyl transformations without any gauging. The gauged Weyl-invariant Wess-Zumino action relevant for our analysis is then equal to
\begin{align} \label{e:WZ}
\delta_\sigma S_{WZ} & = \int \D^4 \bs{x} \sqrt{g} \sigma \left[ ( - e_{TT} + e_{4TT} \phi_0 ) \hat{R}^2 - e_{E44} \phi_0 \hat{E}_4 \right.\nn\\
& \left.\qquad - \: e_{22} \phi_2^2 - e_{44} (\hat{\Box} \phi_0)^2 + 2 e_{2T} \phi_2 \hat{R} + 2 e_{24} \phi_2 \hat{\Box} \phi_0 + 2 e_{4T} \hat{R} \hat{\Box} \phi_0 + \ldots \right].
\end{align}
Due to the homogeneity of each separate term this action satisfies the Wess-Zumino consistency condition trivially (for example $\delta_\sigma (\sqrt{g} \hat{R}^2) = 0$ by construction.) We note however that there may exist a more general solution of the Wess-Zumino consistency condition which involves adding additional terms in the transformation rules (\ref{weyl}) and the action (\ref{e:WZ}). One may analyze this question by expanding the general solution of the Wess-Zumino consistency condition in \cite{Osborn:1991gm, Baume:2014rla} around a fixed point that is a SFT and collecting all terms that are cubic in the sources. We leave such analysis for future work.

In the following sections we will consider higher-point correlation functions as well. When \eqref{e:WZ} is restricted to the metric and the gauge field $C_\mu$ only, one recovers \eqref{e:WZ0}, and hence it contains all terms relevant for computation of scale anomalies in the correlation functions that involve $T$ and $V^\mu$ only.

\subsection{Correlation functions}

In this paper we are mostly interested in correlation functions of the trace of the stress-energy tensor, the virial current and scalar operators of dimensions two and four. These operators, after coupling to background fields, are defined by
\begin{align} \label{e:defT}
& T_{\mu\nu} = \frac{2}{\sqrt{g}} \frac{\delta S}{\delta g^{\mu\nu}}\, , \qquad\qquad T = g^{\mu\nu} T_{\mu\nu}\, , \\
& V^\mu = \frac{1}{\sqrt{g}} \frac{\delta S}{\delta C_\mu}\, , \qquad\qquad \O_\Delta = \frac{1}{\sqrt{g}} \frac{\delta S}{\delta \phi_{d-\Delta}}\, , \label{e:defVO}
\end{align}
Their 1-point functions with sources turned on are then defined as
\begin{align} \label{e:defT2}
& \< T_{\mu\nu} \>_s = - \frac{2}{\sqrt{g}} \frac{\delta W}{\delta g^{\mu\nu}}\, , \qquad\qquad \< T \>_s = g^{\mu\nu} \< T_{\mu\nu} \>_s \, , \\
& \< V^\mu \>_s = - \frac{1}{\sqrt{g}} \frac{\delta W}{\delta C_\mu}\, , \qquad\qquad \< \O_\Delta \>_s = - \frac{1}{\sqrt{g}} \frac{\delta W}{\delta \phi_{d-\Delta}}\, , \label{e:defVOW}
\end{align}
where the subscript $s$ denotes the fact that the operators and their correlation functions are considered with sources turned on. If the subscript is absent, then the correlation function and the operator is considered in the theory with the sources turned off. For example $\< T_{\mu\nu} \> = 0$ since the expectation values of 1-point functions vanish in a SFT, while in general $\< T_{\mu\nu} \>_s \neq 0$ due to the sources.

The theory with sources turned on is Weyl invariant, up to anomalies, when one transforms both the elementary fields and the sources. Let $\Phi$ be the elementary fields (we suppress spacetime and internal indices) transforming  under Weyl transformations as
\begin{equation}
\delta_\sigma \Phi = - \Delta_\Phi \sigma \Phi
\end{equation}
Then (anomalous) Weyl invariance implies
\begin{align}
\delta_\sigma S_{WZ} & = \Big\langle \int \D^{d} \bs{x} \sqrt{g} \left(- \frac{1}{2} T_{\mu \nu} \delta_\sigma g^{\mu \nu}
- V^\mu \delta_\sigma C_{\mu} - \O_\Delta \delta_\sigma \phi_{d-\Delta} - \frac{\delta L}{\delta \Phi} \delta_\sigma \Phi \right) \Big\rangle_s \nonumber \\
&=\Big\langle \int \D^{d} \bs{x} \sqrt{g} \sigma \left(T + \nabla_\mu V^\mu +(d-\Delta) \phi_{d-\Delta} \O_\Delta
+ \Delta_\Phi \Phi \frac{\delta L}{\delta \Phi} \right) \Big\rangle_s
\end{align}
where $L$ is the Lagrangian (including the couplings to sources). This identity should hold for any $\sigma$ and we deduce the scale Ward identity,
\begin{equation} \label{e:scaleWI}
\<T \>_s + \<\nabla_\mu V^\mu \>_s + (d-\Delta) \phi_{d-\Delta} \<\O_\Delta \>_s = \Delta_\Phi \Phi \frac{\delta W}{\delta \Phi} + {\cal A}.
\end{equation}
where ${\cal A}$ is the scale anomaly.

The field equation terms are due to the fact that the elementary fields transform non-trivially under scale transformations. One may remove these terms by defining rescaled elementary fields \cite{Cappelli:1988vw},
\begin{equation}
\Psi = \Phi (\det g)^{\Delta/2d}
\end{equation}
which are Weyl invariant, $\delta_\sigma \Psi =0$. The stress energy tensor will now receive additional contributions
from the variation of the factors of $(\det g)^{\Delta/2d}$ and the Ward identity will not contain any field equations terms which now reads
\begin{equation} \label{e:scaleWI2}
\<T \>_s + \<\nabla_\mu V^\mu \>_s + (d-\Delta) \phi_{d-\Delta} \<\O_\Delta \>_s = {\cal A}.
\end{equation}
The stress energy tensor used in this paper is the one obtained after this redefinition of the elementary fields.

Now, let us take a second Weyl variation of $S_{WZ}$ and then set the sources to zero. Using the fact that the
Wess-Zumino action satisfies the Wess-Zumino condition and that 1-point functions vanish in a SFT we obtain,
\begin{equation}
0=\int \D^4 \bs{x}_1 \sqrt{g} \sigma_1(\bs{x}_1)
\int \D^4 \bs{x}_2 \sqrt{g} \sigma_2(\bs{x}_2) \< (T + \partial_\mu V^\mu)(\bs{x}_1)(T + \partial_\nu V^\nu)(\bs{x}_2) \>
\end{equation}
Since this relation should hold for any $\sigma_1$ and $\sigma_2$ it follows that
\begin{equation}
 \< (T + \partial_\mu V^\mu)(\bs{x}_1)(T + \partial_\nu V^\nu)(\bs{x}_2) \> =0
\end{equation}
and therefore in a unitary SFT,
\begin{equation}
T+\partial_\mu V^\mu=0,
\end{equation}
as an operator equation.
This implies that we can compute correlation function of $T$ either by turning on a source for
$T$ or by turning on a source for $V^\mu$, compute the $V^\mu$ correlators and then act by $\partial_\mu$. This provides consistency conditions that fix some of the semi-local terms that appear in the correlators.

The correlation functions of the trace of the stress-energy tensor may be computed by fixing the background fields to
\begin{equation} \label{e:tau}
g_{\mu\nu} = e^{-2 \tau} \delta_{\mu\nu}, \qquad C_{\mu} = 0, \qquad \phi_{d-\Delta} = 0.
\end{equation}
Indeed, one has
\begin{equation}
2 g^{\mu\nu} \frac{\delta}{\delta g^{\mu\nu}} = \frac{\delta}{\delta \tau}, \qquad\qquad T = e^{d \tau} \frac{\delta S}{\delta \tau}.
\end{equation}
We will also introduce a different parametrisations of the metric given by $\Omega = e^{-\tau}$ and $\Omega = 1 + \varphi$. The correlation functions of the virial current can be computed by fixing the metric $g_{\mu\nu} = \delta_{\mu\nu}$. In terms of the variable $\tau$ the Ricci tensor in arbitrary dimension $d$ reads
\begin{equation}
R[e^{-2 \tau} \delta_{\mu\nu}] = (d-1) e^{2 \tau} \left[ 2 \partial^2 \tau - (d-2) (\partial \tau)^2 \right],
\end{equation}
which allows to express the generating functional in terms of $\tau$.

Since $T_{\mu\nu}$ is a functional of the metric and other sources, the functional derivatives such as $\frac{\delta T_{\mu\nu}}{\delta g^{\rho\sigma}}$ do not vanish in general. The derivative may remain non-zero with sources turned off, allowing us to define under expectation values
\begin{equation} \label{e:deriv}
\< \frac{\delta T}{\delta g^{\rho\sigma}} \ldots \> \stackrel{\text{def}}{=} \< \frac{\delta T}{\delta g^{\rho\sigma}} \ldots \>_{\text{sources} = 0} \qquad\qquad \< \frac{\delta V^\mu}{\delta C_\nu} \ldots \> \stackrel{\text{def}}{=} \< \frac{\delta V^\mu}{\delta C_\nu} \ldots \>_{\text{sources} = 0},
\end{equation}
where `$\ldots$' denote arbitrary operators and similarly for other operators.

In order to keep track of all terms with functional derivatives, we may package them into an interaction action. In four dimensions, up to second order in the dilaton $\tau$ and the gauge field $C_{\mu}$ and to linear order in $\phi_2$ and $\phi_0$ the most general form of the action for any four dimensional reflection positive SFT reads \cite{Luty},
\begin{align} \label{e:Sint}
S_{\text{int}} & = \int \D^4 \bs{x} \left[ \tau T + C_{\mu} V^{\mu} + \phi_2 \O_2 + \phi_0 \O_4 + \ldots \right.\nn\\
& \qquad + \: \tfrac{1}{2} \tau^2 \left( c_T T + c'_2 \partial^2 \O_2 + c_4 \O_4 \right) + \tfrac{1}{2} c_2 (\partial \tau)^2 \O_2 \nn\\
& \qquad \left. + \: \tfrac{1}{2} \tilde{c}_{2} C_\mu C^\mu \O_2 + \ldots \right].
\end{align}
For clarity we consider here the case the SFT contains one operator of dimension two and one operator of dimension four. A generalisation to the case of multiple scalar operators is presented in appendix \ref{sec:multiple}. Other fields are excluded either by the unitarity bounds \eqref{e:bounds}, or can be connected to the terms present by integration by parts. This form of the interaction action is valid in four spacetime dimensions and for reflection-positive theory only. In particular one finds
\begin{align}
\frac{\delta T(\bs{x}_1)}{\delta \tau(\bs{x}_2)} & = 4 T \delta(\bs{x}_1-\bs{x}_2) + \frac{\delta^2 S_{\text{int}}}{\delta \tau(\bs{x}_1) \delta \tau(\bs{x}_2)}, \label{e:dT} \\
\frac{\delta V^{\mu}(\bs{x}_1)}{\delta C_{\nu}(\bs{x}_2)} & = \tilde{c}_2 \delta^{\mu\nu} \O_2 \delta(\bs{x}_1 - \bs{x}_2), \label{e:dV}
\end{align}
where
\begin{align}
\frac{\delta^2 S_{\text{int}}}{\delta \tau(\bs{x}_1) \delta \tau(\bs{x}_2)} = \left[ c_T T + c_2' \partial^2 \O_2 + c_4 \O_4 \right] \delta(\bs{x}_1 - \bs{x}_2) \nn\\
\qquad - \: c_2 \left[ \partial_\mu \O_2 \partial^\mu \delta(\bs{x}_1 - \bs{x}_2) + \O_2 \partial^2 \delta(\bs{x}_1 - \bs{x}_2) \right],
\end{align}
and the derivatives are with respect to $\bs{x}_1$.

\subsection{Scale violations}

Classically, every field in a SFT transforms under dilatations $\bs{x} \mapsto e^\sigma \bs{x}$ in a specific way determined by its scale dimension. For a scalar field $\O$ of dimension $\Delta$ the transformation property reads
\begin{equation}
e^{- \Delta \sigma} \O(e^{- \sigma} \bs{x}) = \O(\bs{x}),
\end{equation}
for any constant $\sigma$. In the quantum theory, however, scale invariance may be violated by logarithmic terms emerging from the renormalisation procedure. The failure of a given $n$-point function to be scale invariant is encoded in the anomaly term $\mathcal{A}_n$. For correlation functions of scalar operators $\O_1, \ldots, \O_n$ of dimensions $\Delta_1, \ldots, \Delta_n$ this implies
\begin{equation}
\mathcal{A}_n(\sigma) = e^{ - \left[ \sum_{j=1}^n \Delta_j - (n-1)d \right] \sigma} \lla \O_1(e^\sigma \bs{p}_1) \ldots \O_n(e^\sigma \bs{p}_n) \rra - \lla \O_1(\bs{p}_1) \ldots \O_n(\bs{p}_n) \rra
\end{equation}
or infinitesimally, to leading order in $\sigma$,
\begin{equation} \label{e:delta}
\hdelta_\sigma \lla \O_1(\bs{p}_1) \ldots \O_n(\bs{p}_n) \rra = \left. \frac{\D}{\D \sigma}  \left(e^{ - \left[ \sum_{j=1}^n \Delta_j - (n-1)d \right] \sigma} \lla \O_1(e^\sigma \bs{p}_1) \ldots \O_n(e^\sigma \bs{p}_n) \rra \right) \right|_{\sigma=0}.
\end{equation}
This variation represents the anomalous contribution. It is equal to the scaling transformation $\delta_\sigma$ up to the classical contribution,
\begin{equation}
\delta_\sigma \lla \O_1(\bs{p}_1) \ldots \O_n(\bs{p}_n) \rra_s = \left[ \hdelta_\sigma + \left( \sum_{j=1}^n \Delta_j - (n-1)d \right) \sigma \right] \lla \O_1(\bs{p}_1) \ldots \O_n(\bs{p}_n) \rra_s.
\end{equation}
This follows from the definition \eqref{e:defVO}, the Fourier transform, and the  simple fact
\begin{equation}
\delta_\sigma \left(\frac{1}{\sqrt{g}} \frac{\delta}{\delta \phi_{d-\Delta}}\right) = \frac{\Delta \sigma}{\sqrt{g}} \frac{\delta}{\delta \phi_{d-\Delta}},
\end{equation}
where $\Delta$ denotes the dimension of the operator sourced by $\phi_{d-\Delta}$. In particular, since $\hdelta_\sigma$ acts on momenta or coordinates rather than sources, the variation $\hdelta_\sigma$ commutes with functional derivatives with respect to the sources,
\begin{equation}
\hdelta_\sigma \< \O_1(\bs{x}_1) \ldots \O_n(\bs{x}_n) \> = \frac{\delta^n}{\delta \phi_1 \ldots \delta \phi_n} \left( \delta_\sigma W \right).
\end{equation}
This implies that the scale violation in the $n$-point function may be calculated by turning the sources off before the variation $\hdelta_\sigma$ is calculated.

The discussion above leads to the conclusion that $\hdelta_\sigma \lla \O_1(\bs{p}_1) \ldots \O_n(\bs{p}_n) \rra = 0$ if anomalies are absent; otherwise logarithmic terms in the $n$-point function appear. In this paper we are mostly interested in correlation functions of the trace of the stress-energy tensor. It follows from the Wess-Zumino action that the only scale violating terms that can appear have a single logarithm of a general form
\begin{equation}
F(\bs{p}_1, \ldots, \bs{p}_n) \log \frac{P(\bs{p}_1, \ldots, \bs{p}_n)}{\mu^2},
\end{equation}
where $F$ is a homogeneous polynomial of degree $\sum_{j=1}^n \Delta_j - (n-1)d$, where $\Delta_j$  are the scaling dimensions of the operators entering the correlator, $P$ is a homogeneous polynomial of degree two and $\mu$ is a renormalisation scale.
Hence from \eqref{e:delta} we find
\begin{equation}
\hdelta_\sigma \lla \O_1(\bs{p}_1) \ldots \O_n(\bs{p}_n) \rra = - \sigma \mu \frac{\D}{\D \mu} \lla \O_1(\bs{p}_1) \ldots \O_n(\bs{p}_n) \rra.
\end{equation}

\section{2-point functions} \label{sec:2pt}

In this section we analyse the structure of 2-point functions in a SFT. We are mostly interested in correlation functions of the stress-energy tensor $T_{\mu\nu}$, the virial current $V^\mu$, and scalar operators $\O_2$ and $\O_4$ of scaling dimension two and four. The form of the 2-point functions is uniquely determined by (the anomalous) scale invariance.
The diagonal part of the matrix of the 2-point functions is
\begin{align}
\lla T(\bs{p}) T(-\bs{p}) \rra & = - e_{TT} p^4 \log p^2 + e_{TT}^{\text{loc}}(\mu) p^4, \label{e:TT} \\
\lla \O_2(\bs{p}) \O_2(-\bs{p}) \rra & = - e_{22} \log p^2 + e_{22}^{\text{loc}}(\mu), \label{e:22} \\
\lla \O_4(\bs{p}) \O_4(-\bs{p}) \rra & = - e_{44} p^4 \log p^4 + e_{44}^{\text{loc}}(\mu) p^4, \label{e:44}
\end{align}
while the off-diagonal part reads
\begin{align}
\lla T(\bs{p}) \O_2(-\bs{p}) \rra & = e_{2T} p^2 \log p^2 + e_{2T}^{\text{loc}}(\mu) p^2, \label{e:2T} \\
\lla T(\bs{p}) \O_4(-\bs{p}) \rra & = e_{4T} p^4 \log p^2 + e_{4T}^{\text{loc}}(\mu) p^4, \label{e:4T} \\
\lla \O_2(\bs{p}) \O_4(-\bs{p}) \rra & = e_{24} p^2 \log p^2 + e_{24}^{\text{loc}}(\mu) p^2. \label{e:24}
\end{align}
The normalisation constants defined here correspond to the constants featuring in the Wess-Zumino action \eqref{e:WZ}. Indeed, by taking two derivatives of the generating functional with respect to the dilaton $\tau$, \eqref{e:tau}, one finds,
\begin{equation}
\< T(\bs{x}_1) T(\bs{x}_2) \> = \frac{\delta^2 W}{\delta \tau(\bs{x}_1) \delta \tau(\bs{x}_2)}.
\end{equation}
Note that such a simple expression holds only after the sources are turned off.

The local, scheme dependent part of each correlation function can be adjusted by means of finite local counterterms. In a reflection positive QFT the matrix of the normalisation constants must be non-negative. Furthermore, if one of the eigenvalues vanishes, then the corresponding operator is null. In particular if $e_{TT} = 0$ (after we set to zero the off-diagonal terms as discussed in the next subsection), then $T = 0$ and the scale invariant theory is fully conformally invariant.

Furthermore, scale invariance requires that the source $C_\mu$ for the virial current appears in a correlated way with the metric.
Therefore, the normalisation constants of the 2-point functions of the virial current with other operators are related with the corresponding correlation functions of $T$. In particular,
\begin{align}
\lla \O_2(\bs{p}) V^\mu(-\bs{p}) \rra & = - \I e_{2T} p^\mu \log p^2 + \I e_{2V}^{\text{loc}} p^\mu, \\
\lla T(\bs{p}) V^\mu(-\bs{p}) \rra & = \I e_{TT} p^2 p^\mu \log p^2 - \I e_{TV}^{\text{loc}}(\mu) p^2 p^\mu, \\
\lla \O_4(\bs{p}) V^\mu(-\bs{p}) \rra & = \I e_{4T} p^2 p^\mu \log p^2 - \I e_{4T}^{\text{loc}}(\mu) p^2 p^\mu, \\
\lla V^{\mu}(\bs{p}) V^\nu(-\bs{p}) \rra & = - e_{TT} p^\mu p^\nu \log p^2 + e_{TT}^{\text{loc}}(\mu) p^\mu p^\nu + \text{transverse part}.
\end{align}
The transverse part of the 2-point function of the virial current is proportional to $\pi^{\mu\nu} = \delta^{\mu\nu} - p^\mu p^\nu / p^2$ and is not relevant for our discussion (which involve checking the implications
of the relation $T = - \partial_\mu V^\mu$) because $p_\mu \pi^{\mu\nu} = 0$.
Furthermore in this paper we are interested in calculating scale violations in the correlation functions so from now on we will neglect the local parts.

\subsection{Improvement term} \label{sec:improv}

Consider a SFT with $e_{TT} > 0$ and at least one scalar operator $\O_2$ of dimension two with $e_{22} > 0$ in the interaction action \eqref{e:Sint}. The case of multiple scalar operators of dimensions two and four is discussed in appendix \ref{sec:multiple}. The improvement term
\begin{equation} \label{e:imp0}
\Delta S = \frac{\xi}{6} \int \D^4 \bs{x} \sqrt{g} R \O_2
\end{equation}
can be added to the action in such a way that the improved stress-energy tensor becomes traceless. The improvement term does not alter the charges associated with the stress-energy tensor and modifies the correlation functions only locally. The trace of the improved stress-energy tensor reads
\begin{equation}
T \mapsto T_{\text{imp}} = T + \xi \partial^2 \O_2.
\end{equation}
Clearly, if $T = c \partial^2 \O_2$ for some constant $c$, then for $\xi = -c$ the trace of the improved stress-energy tensor vanishes. Otherwise, the 2-point function of the improved stress-energy tensor reads
\begin{equation}
\lla T_{\text{imp}}(\bs{p}) T_{\text{imp}}(-\bs{p}) \rra = \lla T(\bs{p}) T(-\bs{p}) \rra - 2 \xi p^2 \lla T(\bs{p}) \O_2(- \bs{p}) \rra + \xi^2 p^4 \lla \O_2(\bs{p}) \O_2(-\bs{p}) \rra.
\end{equation}
By using \eqref{e:TT} - \eqref{e:24} we find that the 2-point function of the trace of the improved stress-energy tensor vanishes if $\xi$ satisfies the equation
\begin{equation}
e_{TT} + 2 \xi e_{2T} + \xi^2 e_{22} = 0.
\end{equation}
A solution exists if and only if
\begin{equation} \label{imp1}
e_{2T}^2 - e_{TT} e_{22} \geq 0.
\end{equation}
Now we will show that such a condition can hold in a reflection positive theory if and only if $T$ is proportional to $\partial^2 \O_2$. To do it, consider the state $| \Psi \>$ defined as
\begin{equation} \label{e:state}
| \Psi \> = \alpha T(x) | 0 \> + \beta \partial^2 \O_2(x) | 0 \>,
\end{equation}
where $\alpha$ and $\beta$ are arbitrary complex numbers. Reflection positivity implies that the norm $\< \Psi | \Psi \>$ must be non-negative. Poincar\'{e} invariance, together with the fact that complex conjugation in an Euclidean setting corresponds to time reversal leads to the conclusion that
\begin{equation} \label{imp2}
e_{2T}^2 - e_{TT} e_{22} \leq 0,
\end{equation}
in any reflection positive QFT. Therefore (\ref{imp1}) and (\ref{imp2}) are compatible (and then the improvement term exists) if and only if  $e_{2T}^2 - e_{TT} e_{22} = 0$. In this case \eqref{e:state} implies that states $T(x) | 0 \>$ and $\partial^2 \O_2(x) | 0 \>$ are linearly dependent, and so $T = c \partial^2 \O_2$ for some $c$, and the theory is conformal. We therefore assume from now on that \begin{equation} \label{ineq}
e_{2T}^2 - e_{TT} e_{22} < 0.
\end{equation}

\subsection{Diagonalising the 2-point functions} \label{sec:imp}

We just argued that when (\ref{ineq}) holds one cannot improve $T$. However, one can use the improvement terms in order to fix the off-diagonal 2-point functions \eqref{e:2T} - \eqref{e:24} to zero. This is motivated by conformal field theories, where such correlation functions vanish. To do it, consider a general form of the improvement term
\begin{equation} \label{e:imp}
\Delta S = \int \sqrt{g} \left[ \xi \O_2 \Sigma + \xi' \O_2 \hat{\Box} \phi_0 + \xi'' \phi_0 T \right].
\end{equation}
The first term contains the previously considered improvement term \eqref{e:imp0}, now written in a Weyl invariant way. The existence of the improvement terms follows from the fact that the dimension of the source for the operator $\O_2$ is the same as for the operator itself. Therefore, one can introduce the improvement terms simply by exchanging one $\phi_2$ in favour of $\O_2$ in the Wess-Zumino action \eqref{e:WZ}. The improvement term modifies the operators in the theory as follows,
\begin{align}
T_{\text{imp}} & = T + \xi \partial^2 \O_2, \\
V^{\mu}_{\text{imp}} & = V^{\mu} - \xi \partial_\mu \O_2, \\
\O_4^{\text{imp}} & = \O_4 + \xi' \partial^2 \O_2 + \xi'' T.
\end{align}
The off-diagonal 2-point functions of the improved operators vanish if the following system has a solution
\begin{align}
0 & = e_{2T} + \xi e_{TT}, \label{e:Timp} \\
0 & = e_{24} + \xi' e_{22} + \xi'' e_{2T}, \\
0 & = e_{4T} - e_{2T} (\xi' + \xi \xi'') - \xi e_{24} - \xi'' e_{TT} - \xi \xi' e_{22}. \label{e:O4imp}
\end{align}
The solution exists if $e_{TT} e_{22} \neq e_{2T}^2$ and reads
\begin{align}
\xi & = - \frac{e_{2T}}{e_{TT}}, \\
\xi' & = - \frac{e_{2T} e_{4T} + e_{24} e_{TT}}{e_{22} e_{TT} - e_{2T}^2}, \\
\xi'' & =  \frac{e_{24} e_{2T} + e_{22} e_{4T}}{e_{22} e_{TT} - e_{2T}^2}.
\end{align}
As argued in the previous section the denominators of these expressions are non-vanishing in any reflection positive QFT containing an operator of dimension two. Therefore we have shown that in such a case one can always add an improvement and from now on assume that
\begin{equation} \label{e:eare0}
e_{2T} = e_{24} = e_{4T} = 0.
\end{equation}
If the operator of dimension two is absent in the theory, then already $e_{22} = e_{24} = e_{2T} = 0$. In this case the theory can be improved to $e_{4T} = 0$ by a simple shift given by $\xi''$. Therefore, from now on, we assume that \eqref{e:eare0} holds.

Furthermore, if multiple scalar operators of dimension two and four are present in the theory, one can generalise the procedure and show that all off-diagonal correlation functions vanish. We discuss the general case in appendix \ref{sec:multiple}. For clarity, in the main text of the paper we assume that the theory contains at most one operator of dimension two and four (other than the trace of the stress-energy tensor).

\section{3-point functions} \label{sec:3pt}

In this section we will analyse the structure of scale violating terms in the 3-point functions $\< TTT \>$, $\< TT \O_2 \>$ and $\<TT \O_4 \>$. We will compare the scale violations following from the Wess-Zumino action \eqref{e:WZ} with possible scale violation in the analogous correlation functions involving the virial current. Then the two expressions can be compared by means of the relation $T = - \partial_\mu V^\mu$.

\subsection{\texorpdfstring{$\<TTT\>$}{<TTT>}} \label{sec:TTT}

The correlation functions of the stress-energy tensor are defined as correlation functions of the operator defined in \eqref{e:defT}. In particular, the relation between the actual 3-point function of the trace of the stress-energy tensor and the triple functional derivative of the generating functional with respect to the metric involves semi-local terms, \textit{i.e.}, terms that in position space are supported on the set of coincident points. In all expressions we can omit contributions from 1-point functions, since after turning off the sources they vanish. However, we will carefully account for all 2-point functions including terms with functional derivatives.

Using the parametrisation of the metric as $g_{\mu\nu} = e^{-2 \tau} \delta_{\mu\nu}$ one finds,
\begin{align} \label{e:TTT1}
& \left( - e^{d \tau(\bs{x}_1)} \frac{\delta}{\delta \tau(\bs{x}_1)} \right) \left( - e^{d \tau(\bs{x}_2)} \frac{\delta}{\delta \tau(\bs{x}_2)} \right) \left( - e^{d \tau(\bs{x}_3)} \frac{\delta}{\delta \tau(\bs{x}_3)} \right) W = \\
& = \< T(\bs{x}_1) T(\bs{x}_2) T(\bs{x}_3) \>_s - \left[ e^{d \tau(\bs{x}_1)} \< \frac{\delta T(\bs{x}_2)}{\delta \tau(\bs{x}_1)} T(\bs{x}_3) \>_s + 2 \text{ permutations} \right]\nn\\
& \qquad\qquad + 1\text{-point functions}.
\end{align}
Carrying out the derivatives on the left hand side, using \eqref{e:dT} and then turning off the sources we find
\begin{align} \label{e:TTT2}
& \< T(\bs{x}_1) T(\bs{x}_2) T(\bs{x}_3) \> = - \frac{\delta^3 W}{\delta \tau(\bs{x}_1) \delta \tau(\bs{x}_2) \delta \tau(\bs{x}_3)} \\
& \qquad + \: \left[ \< \frac{\delta^2 S_{\text{int}}}{\delta \tau(\bs{x}_1) \delta \tau(\bs{x}_2)} T(\bs{x}_3) \> + \< \frac{\delta^2 S_{\text{int}}}{\delta \tau(\bs{x}_2) \delta \tau(\bs{x}_3)} T(\bs{x}_1) \> + \< \frac{\delta^2 S_{\text{int}}}{\delta \tau(\bs{x}_3) \delta \tau(\bs{x}_1)} T(\bs{x}_2) \> \right], \nn
\end{align}
where the interaction action $S_{\text{int}}$ is given in \eqref{e:Sint}. Due to the improvement terms introduced in section \ref{sec:imp}, the only contribution to the 2-point functions comes from the $\tfrac{1}{2} \tau^2 c_T T$ term in the interaction action. Therefore,
\begin{align} \label{e:TTT3}
& \< T(\bs{x}_1) T(\bs{x}_2) T(\bs{x}_3) \> = - \frac{\delta^3 W}{\delta \tau(\bs{x}_1) \delta \tau(\bs{x}_2) \delta \tau(\bs{x}_3)} \nn\\
& \qquad + \: c_T \left[ \delta(\bs{x}_1 - \bs{x}_2) \< T(\bs{x}_2) T(\bs{x}_3) \> + \delta(\bs{x}_2 - \bs{x}_3) \< T(\bs{x}_3) T(\bs{x}_1) \> + \delta(\bs{x}_3 - \bs{x}_1) \< T(\bs{x}_1) T(\bs{x}_2) \> \right].
\end{align}
The scale violation in the 3-point function can be computed from the Wess-Zumino action and the expressions for 2-point functions. The result is
\begin{align} \label{e:anomal1}
& \hdelta_\sigma \lla T(\bs{p}_1) T(\bs{p}_2) T(\bs{p}_3) \rra =  2 \sigma e_{TT} J^2 - 2 \sigma c_T e_{TT} (p_1^4 + p_2^4 + p_3^4),
\end{align}
where $J^2$ is defined in \eqref{e:J2}.

Next we can compare the scale violation \eqref{e:anomal1} with the scale violation following from the virial current. We follow the same procedure. First we expand the triple derivative of the generating functional with respect to the source $C_\mu$,
\begin{align}
& - \frac{\delta^3 W}{\delta C_{\mu_1}(\bs{x}_1) \delta C_{\mu_2}(\bs{x}_2) \delta C_{\mu_3}(\bs{x}_3)} = \< V^{\mu_1}(\bs{x}_1) V^{\mu_2}(\bs{x}_2) V^{\mu_3}(\bs{x}_3) \>_s \nn\\
& \qquad - \: \left[ \< \frac{\delta V^{\mu_1}(\bs{x}_1)}{\delta C_{\mu_3}(\bs{x}_3)} V^{\mu_2}(\bs{x}_2) \>_s + 2 \text{ cycl. perm.} \right] + \text{1-point functions}.
\end{align}
The functional derivatives can be read off from the action \eqref{e:Sint} using \eqref{e:dV} and the scale violation can be calculated by means of the Wess-Zumino action. In total one finds
\begin{equation}
\I \hdelta_\sigma \lla V^{\mu_1}(\bs{p}_1) V^{\mu_2}(\bs{p}_2) V^{\mu_3}(\bs{p}_3) \rra = - 4 e_{TT} \sigma \left[ p_1^{\mu_1} \delta^{\mu_2 \mu_3} + p_2^{\mu_2} \delta^{\mu_1 \mu_3} + p_3^{\mu_3} \delta^{\mu_1 \mu_2} \right].
\end{equation}
Using $T = - \partial_\mu V^\mu$ we find
\begin{equation} \label{e:TTT}
\hdelta_\sigma \lla T(\bs{p}_1) T(\bs{p}_2) T(\bs{p}_3) \rra = 2 e_{TT} \sigma J^2.
\end{equation}

By comparison of \eqref{e:anomal1} and \eqref{e:TTT} we find that either $e_{TT} = 0$ and the theory is conformal, or $c_T = 0$. Therefore, from now on we assume $c_T = 0$.

Finally, we note that the most general form of the scale violation in the 3-point function of the virial current can be derived using Lorentz and scale invariance and the fact that anomalies are local. We present this alternative derivation in appendix \ref{sec:p2}.

\subsection{\texorpdfstring{$\<TT \O_2\>$}{<TTO[2]>}} \label{sec:TT2}

In the forthcoming analysis of the 4-point functions we will also need to explore the consequences of the relation between $\< T T \O_2 \>$ and $\< V^{\mu} V^{\nu} \O_2 \>$. Following the procedure described in the previous section the scale violation in the virial current leads to
\begin{equation}
\hdelta_\sigma \lla V^{\mu}(\bs{p}_1) V^{\nu}(\bs{p}_2) \O_2(\bs{p}_3) \rra = - 2 \sigma \tilde{c}_2 e_{22} \delta^{\mu\nu},
\end{equation}
which results in
\begin{equation} \label{e:TT2}
\hdelta_\sigma \lla T(\bs{p}_1) T(\bs{p}_2) \O_2(\bs{p}_3) \rra = 2 \sigma e_{22} \tilde{c}_2 \bs{p}_1 \cdot \bs{p}_2.
\end{equation}

On the other hand the scale violation in the correlator involving the trace of the stress-energy tensor can be computed directly from the Wess-Zumino action. In this way one finds
\begin{equation} \label{e:TT2_2}
\hdelta_\sigma \lla T(\bs{p}_1) T(\bs{p}_2) \O_2(\bs{p}_3) \rra = 2 \sigma e_{22} ( c_2 \bs{p}_1 \cdot \bs{p}_2 - c_2' p_3^2 ).
\end{equation}
Using the relation $\bs{p}_1 \cdot \bs{p}_2 = \tfrac{1}{2} (p_3^2 - p_1^2 - p_2^2)$ we can rewrite the correlation function in terms of the three independent magnitudes of momenta $p_1, p_2$ and $p_3$. It follows that \eqref{e:TT2} and \eqref{e:TT2_2} agree only if
\begin{equation}
c_2' = 0, \qquad\qquad c_2 = \tilde{c}_2.
\end{equation}

\subsection{\texorpdfstring{$\<TT \O_4\>$}{<TTO[4]>}} \label{sec:TT4}

In this subsection we compute $\<TT \O_4\>$ and $\< V^{\mu} V^{\nu} \O_4 \>$, where $\O_4$ denotes the operator of dimension four, and find that the coefficient $c_4$ in \eqref{e:Sint} must vanish, $c_4 = 0$. This 3-point function receives a contribution from the term $e_{4TT} \int \sqrt{g} \phi_0 \Sigma^2$ in the Wess-Zumino action \eqref{e:WZ}.

Taking three derivatives of the generating functional with respect to appropriate sources we find
\begin{align}
& \< T(\bs{x}_1) T(\bs{x}_2) \O_4(\bs{x}_3) \> = - \frac{\delta^3 W}{\delta \tau(\bs{x}_1) \delta \tau(\bs{x}_2) \delta \phi_0(\bs{x}_3)} \nn\\
& \qquad + \: \left[ \< \frac{\delta T(\bs{x}_1)}{\delta \phi_0(\bs{x}_3)} T(\bs{x}_2) \> + \< \frac{\delta T(\bs{x}_2)}{\delta \phi_0(\bs{x}_3)} T(\bs{x}_1) \> + \< \frac{\delta T(\bs{x}_1)}{\delta \tau(\bs{x}_2)} \O_4(\bs{x}_2) \> \right].
\end{align}
The operator $\delta T/\delta \phi_0$ has dimension four and hence can be written in a basis of such operators. However, the only operator in such expansion that can produce a non-zero answer is the trace of stress the energy tensor (since we arranged for all off-diagonal 2-point functions to be equal to zero). Therefore, using the Wess-Zumino action, one finds
\begin{equation} \label{e:TT4a}
\hdelta_\sigma \lla T(\bs{p}_1) T(\bs{p}_2) \O_4(\bs{p}_3) \rra = - 2 \sigma e_{4TT} p_1^2 p_2^2 - 2 \sigma C e_{TT} (p_1^4 + p_2^4) - 4 \sigma c_{E44} J^2 - 2 \sigma c_4 e_{44} p_3^4,
\end{equation}
where $C$ is some numerical constant.

On the other hand, a similar calculation can be carried out using the virial current. By taking three functional derivatives one finds
\begin{align}
&\< V^\mu(\bs{x}_1) V^\nu(\bs{x}_2) \O_4(\bs{x}_3) \> = - \frac{\delta^3 W}{\delta C_\mu(\bs{x}_1) \delta C_\nu(\bs{x}_2) \delta \phi_0(\bs{x}_3)} \nn\\
& \qquad + \: \left[ \< \frac{\delta V^\mu(\bs{x}_1)}{\delta \phi_0(\bs{x}_3)} V^\nu(\bs{x}_2) \> + \< \frac{\delta V^\nu(\bs{x}_2)}{\delta \phi_0(\bs{x}_3)} V^\mu(\bs{x}_1) \> + \< \frac{\delta V^\mu(\bs{x}_1)}{\delta C_\nu(\bs{x}_2)} \O_4(\bs{x}_2) \> \right].
\end{align}
The functional derivative of the virial current with respect to $\phi_0$ is an operator of dimension three and hence we can write its most general form
\begin{equation}
\frac{\delta V^\mu(\bs{x})}{\delta \phi_0(\bs{y})} = \delta(\bs{x} - \bs{y}) \left[ \sum_k a_k j_k^\mu + b \partial^\mu \O_2 \right] + c \O_2 \partial^\mu \delta(\bs{x} - \bs{y}),
\end{equation}
for some constants $a_k, b, c$, where $j_k^\mu$ is a set of currents of dimension $3$ including possibly $V^\mu$. This leads to
\begin{equation}
\hdelta_\sigma  \lla T(\bs{p}_1) T(\bs{p}_2) \O_4(\bs{p}_3) \rra = - 2 \sigma e_{4TT} p_1^2 p_2^2 - 4 \sigma c_{E44} J^2 - 2 \sigma e_{Vj} (p_1^4 + p_2^4),
\end{equation}
where $e_{Vj}$ is a total normalisation constant following from the sum of all 2-point functions $\< j_k^\mu V^\nu \>$. Since the coefficient of $p_3^4$ vanishes in this expression, by comparing with \eqref{e:TT4a} we obtain either $e_{44} = 0$ or $c_4 = 0$.
If $e_{44}=0$, the operator $\O_4$ is null and may be set to zero. Otherwise, $c_4=0$. In any case,
in the computations to follow only the product $e_{44} c_4$ appears and this vanishes in both cases.

This cancellation can be explained as follows. Since $T = g^{\mu\nu} T_{\mu\nu}$, where $T_{\mu\nu}$ is the full stress-energy tensor with sources turned on, the only term containing the metric and the operator $\O_4$ in $T_{\mu\nu}$ can be $g_{\mu\nu} \O_4$. However, now $T = 4 \O_4$ and the functional derivative with respect to the metric vanishes, hence $c_4 = 0$.

\section{4-point functions} \label{sec:4pt}

In this section we follow the same procedure as before applied to connected 4-point functions. We will compare the scale violations in the 4-point function of the trace of the stress-energy tensor with the scale violations in the virial current.

In the previous section we found a series of conditions relating various coefficients in the functional derivative terms and the interaction action \eqref{e:Sint}. With all these conditions, the action reads now
\begin{align} \label{e:Sint2}
S_{\text{int}} & = \int \D^4 \bs{x} \left[ \tau T + C_{\mu} V^{\mu} + \phi_2 \O_2 + \phi_0 \O_4 + \ldots \right.\nn\\
& \qquad \left. + \: \tfrac{1}{2} c_2 (\partial \tau)^2 \O_2 + \tfrac{1}{2} c_2 C_\mu C^\mu \O_2 + \ldots \right]
\end{align}
with an undetermined value of $c_2$.

\subsection{\texorpdfstring{$\< V^{\mu_1} V^{\mu_2} V^{\mu_3} V^{\mu_4} \>$}{<VVVV>}} \label{sec:VVVV}

We start by computing the scale violating terms in $\< V^{\mu_1} V^{\mu_2} V^{\mu_3} V^{\mu_4} \>$. Since the total dimension of this correlation function in momentum space equals zero, the scale violating terms must be proportional to the unique symmetric tensor of dimension zero,
\begin{equation} \label{e:Stensor}
S^{\mu_1 \mu_2 \mu_3 \mu_4} = \delta^{\mu_1 \mu_2} \delta^{\mu_3 \mu_4} + \delta^{\mu_1 \mu_3} \delta^{\mu_2 \mu_4} + \delta^{\mu_1 \mu_4} \delta^{\mu_2 \mu_3}.
\end{equation}

As in section \ref{sec:TTT}, we first obtain the relation between derivatives of the generating functional and the 4-point function:
\begin{align}
& \< V^{\mu_1}(\bs{x}_1) V^{\mu_2}(\bs{x}_2) V^{\mu_3}(\bs{x}_3) V^{\mu_4}(\bs{x}_4) \> = \frac{\delta^4 W}{\delta C_{\mu_1}(\bs{x}_1) \delta C_{\mu_2}(\bs{x}_2) \delta C_{\mu_3}(\bs{x}_3) \delta C_{\mu_4}(\bs{x}_4)} \nn\\
& \qquad + \: \left[ \< \frac{\delta V^{\mu_1}(\bs{x}_1)}{\delta C_{\mu_2}(\bs{x}_2)} V^{\mu_3}(\bs{x}_3) V^{\mu_4}(\bs{x}_4) \> + 5 \text{ permutations} \right] \nn\\
& \qquad - \: \left[ \< \frac{\delta V^{\mu_1}(\bs{x}_1)}{\delta C_{\mu_2}(\bs{x}_2)} \frac{\delta V^{\mu_3}(\bs{x}_3)}{\delta C_{\mu_4}(\bs{x}_4)} \> +  \< \frac{\delta V^{\mu_1}(\bs{x}_1)}{\delta C_{\mu_3}(\bs{x}_3)} \frac{\delta V^{\mu_2}(\bs{x}_2)}{\delta C_{\mu_4}(\bs{x}_4)} \> + \< \frac{\delta V^{\mu_1}(\bs{x}_1)}{\delta C_{\mu_4}(\bs{x}_4)} \frac{\delta V^{\mu_2}(\bs{x}_2)}{\delta C_{\mu_3}(\bs{x}_3)} \> \right] \nn\\
& \qquad - \: \left[ \< \frac{\delta^2 V^{\mu_1}(\bs{x}_1)}{\delta C_{\mu_2}(\bs{x}_2) \delta C_{\mu_3}(\bs{x}_3)} V^{\mu_4}(\bs{x}_4) \> + 3 \text{ permutations} \right].
\end{align}
The second functional derivative of the virial current with respect to $C_\mu$ has dimension one and hence it vanishes. By using equations \eqref{e:Sint2} and \eqref{e:TT2} and the Wess-Zumino action one finds the scale violation
\begin{equation}
\hdelta_\sigma \lla V^{\mu_1}(\bs{p}_1) V^{\mu_2}(\bs{p}_2) V^{\mu_3}(\bs{p}_3) V^{\mu_4}(\bs{p}_4) \rra = - 8 \sigma S^{\mu_1 \mu_2 \mu_3 \mu_4} \left( e_{TT} + \tfrac{1}{4} c_2^2 e_{22} \right).
\end{equation}
Therefore the scale violation in the 4-point function of the trace of the stress-energy tensor reads
\begin{align} \label{e:TTTTfromV}
& \hdelta_\sigma \lla T(\bs{p}_1) T(\bs{p}_2) T(\bs{p}_3) T(\bs{p}_4) \rra = - 8 \sigma \left( e_{TT} + \tfrac{1}{4} c_2^2 e_{22} \right) \times \nn\\
& \qquad \times \left[ (\bs{p}_1 \cdot \bs{p}_2) (\bs{p}_3 \cdot \bs{p}_4) + (\bs{p}_1 \cdot \bs{p}_3) (\bs{p}_2 \cdot \bs{p}_4) + (\bs{p}_1 \cdot \bs{p}_4) (\bs{p}_2 \cdot \bs{p}_3) \right].
\end{align}
Note that this is an exact result and it is non-vanishing even in the forward scattering limit \eqref{e:fwd},
\begin{equation}
\left. \hdelta_\sigma \lla T(\bs{p}_1) T(\bs{p}_2) T(\bs{p}_3) T(\bs{p}_4) \rra \right|_{p_j^2 = 0, t = 0} = - 4 \sigma s^2 \left( e_{TT} + \tfrac{1}{4} c_2^2 e_{22} \right).
\end{equation}
This suggests that $T$ may be a non-trivial operator in a SFT.

\subsection{\texorpdfstring{$\< T T T T \>$}{<TTTT>}} \label{sec:TTTT}

In this subsection we carry out the computation of the scale violating terms in the 4-point function of the trace of the stress-energy tensor directly from the Wess-Zumino action. The calculation is long but otherwise straightforward. As in previous sections, we start by evaluating four functional derivatives of the generating functional with respect to the dilaton. After turning off the sources one finds
\begin{align}
& \left( - e^{d \tau(\bs{x}_1)} \frac{\delta}{\delta \tau(\bs{x}_1)} \right) \left( - e^{d \tau(\bs{x}_2)} \frac{\delta}{\delta \tau(\bs{x}_2)} \right) \left( - e^{d \tau(\bs{x}_3)} \frac{\delta}{\delta \tau(\bs{x}_3)} \right) \left( - e^{d \tau(\bs{x}_4)} \frac{\delta}{\delta \tau(\bs{x}_4)} \right) W = \nn\\
& = \< T(\bs{x}_1) T(\bs{x}_2) T(\bs{x}_3) T(\bs{x}_4) \> - \left[ \< \frac{\delta T(\bs{x}_1)}{\delta \tau(\bs{x}_2)} T(\bs{x}_3) T(\bs{x}_4) \> + 5 \text{ permutations} \right] \nn\\
& \qquad + \: \left[ \< \frac{\delta T(\bs{x}_1)}{\delta \tau(\bs{x}_2)} \frac{\delta T(\bs{x}_3)}{\delta \tau(\bs{x}_4)} \> + \< \frac{\delta T(\bs{x}_1)}{\delta \tau(\bs{x}_3)} \frac{\delta T(\bs{x}_2)}{\delta \tau(\bs{x}_4)} \> + \< \frac{\delta T(\bs{x}_1)}{\delta \tau(\bs{x}_4)} \frac{\delta T(\bs{x}_2)}{\delta \tau(\bs{x}_3)} \> \right] \nn\\
& \qquad + \: \left[ \< \frac{\delta^2 T(\bs{x}_1)}{\delta \tau(\bs{x}_2) \delta \tau(\bs{x}_3)} T(\bs{x}_4) \> + 3 \text{ permutations} \right] \nn\\
& \qquad + \: d \left[ \delta(\bs{x}_1 - \bs{x}_2) \< \frac{\delta T(\bs{x}_2)}{\delta \tau(\bs{x}_3)} T(\bs{x}_4) \> + 3 \text{ permutations} \right].
\end{align}
The left hand side can be expanded and the result is then expressed in terms of functional derivatives of the interaction action \eqref{e:Sint2} using \eqref{e:dT}. Most terms cancel and one finds
\begin{align} \label{e:TTTTtowork}
& \< T(\bs{x}_1) T(\bs{x}_2) T(\bs{x}_3) T(\bs{x}_4) \> = \frac{\delta^4 W}{\delta \tau(\bs{x}_1) \delta \tau(\bs{x}_2) \delta \tau(\bs{x}_3) \delta \tau(\bs{x}_4)} \nn\\
& \qquad + \: \left[ \< \frac{\delta^2 S_{\text{int}}}{\delta \tau(\bs{x}_1) \delta \tau(\bs{x}_2)} T(\bs{x}_3) T(\bs{x}_4) \> + 5 \text{ permutations} \right] \nn\\
& \qquad - \: \left[ \< \frac{\delta^2 S_{\text{int}}}{\delta \tau(\bs{x}_1) \delta \tau(\bs{x}_2)} \frac{\delta^2 S_{\text{int}}}{\delta \tau(\bs{x}_3) \delta \tau(\bs{x}_4)} \> + 2 \text{ permutations} \right] \nn\\
& \qquad - \: \left[ \< \frac{\delta^3 S_{\text{int}}}{\delta \tau(\bs{x}_1) \delta \tau(\bs{x}_2) \delta \tau(\bs{x}_3)} T(\bs{x}_4) \> + 3 \text{ permutations} \right].
\end{align}
The scale violation of the first term on the right hand side follows from the Wess-Zumino action and reads
\begin{align} \label{e:dZinTTTT}
& \hdelta_\sigma \frac{\delta^4 S_{WZ}}{\delta \tau(\bs{p}_1) \delta \tau(\bs{p}_2) \delta \tau(\bs{p}_3) \delta \tau(\bs{p}_4)} = - 8 \sigma e_{TT} \times \nn\\
& \qquad\qquad \times \left[ (\bs{p}_1 \cdot \bs{p}_2) (\bs{p}_3 \cdot \bs{p}_4) + (\bs{p}_1 \cdot \bs{p}_3) (\bs{p}_2 \cdot \bs{p}_4) + (\bs{p}_1 \cdot \bs{p}_4) (\bs{p}_2 \cdot \bs{p}_3) \right].
\end{align}
As we see that the scale violation from the Wess-Zumino action already matches the first term in \eqref{e:TTTTfromV}.

In order to proceed  we need to first analyse the term with three $\tau$-derivatives. This term receives contributions from terms cubic in the dilaton in the interaction action and these terms are not listed in \eqref{e:Sint2}. However, since the third derivative of the action appears in \eqref{e:TTTTtowork} under the expectation value with the trace of the stress-energy tensor, the only relevant term in the interaction action is $\int \tfrac{1}{6} \tau^3 c_T^{(3)} T$. By taking three derivatives one finds
\begin{equation}
\hdelta_\sigma \lla \frac{\delta^3 S_{\text{int}}}{\delta \tau(\bs{p}_1) \delta \tau(\bs{p}_2) \delta \tau(\bs{p}_3)} T(\bs{p}_4) \rra = - 2 \sigma c_T^{(3)} e_{TT} p_4^4.
\end{equation}
The remaining computations are straightforward. The result reads
\begin{align} \label{e:TTTTfromT}
& \hdelta_\sigma \lla T(\bs{p}_1) T(\bs{p}_2) T(\bs{p}_3) T(\bs{p}_4) \rra = - 8 \sigma \left( e_{TT} + \tfrac{1}{4} c_2^2 e_{22} \right) \times \nn\\
& \qquad\qquad \times \left[ (\bs{p}_1 \cdot \bs{p}_2) (\bs{p}_3 \cdot \bs{p}_4) + (\bs{p}_1 \cdot \bs{p}_3) (\bs{p}_2 \cdot \bs{p}_4) + (\bs{p}_1 \cdot \bs{p}_4) (\bs{p}_2 \cdot \bs{p}_3) \right] \nn\\
& \qquad - \: 2 \sigma c_T^{(3)} e_{TT} \left( p_1^4 + p_2^4 + p_3^4 + p_4^4 \right).
\end{align}
By comparing with \eqref{e:TTTTfromV} we see that the leading term involving $e_{TT}$ and $e_{22}$ matches exactly and then one is forced to take $c_T^{(3)} = 0$.

To summarize, the anomaly in the 4-point function is given by \eqref{e:TTTTfromV}  and the following relations among
the second order coefficients in the interaction action \eqref{e:Sint} hold,
\begin{equation} \label{e:allvanish}
c_T = \tilde{c}_2 = c_4 = 0, \qquad\qquad \tilde{c}_2 = c_2.
\end{equation}

Note that while the total scale violation of any correlation function is invariant under parametrisations of the metric, the source of the various contributions do depend on such parametrisations. With $g_{\mu\nu} = e^{-2 \tau} \delta_{\mu\nu}$ we have found a non-zero contribution from the Wess-Zumion action given by \eqref{e:dZinTTTT}. Such a contribution does not vanish in the forward scattering limit \eqref{e:fwd}. On the other hand, if one parametrises the metric as $g_{\mu\nu} = \Omega^2 \delta_{\mu\nu}$, then one finds
\begin{equation}
\frac{\delta^4}{\delta \Omega(\bs{p}_1) \delta \Omega(\bs{p}_2) \delta \Omega(\bs{p}_3) \delta \Omega(\bs{p}_4)} (\delta_\sigma S_{WZ}) = - 12 \sigma e_{TT} \sum_{1 \leq i < j \leq 4} p_i^2 p_j^2.
\end{equation}
In this parametrisation of the metric the contribution from the Wess-Zumino action vanishes in the forward scattering limit with $p_j^2 = 0$, $j=1,2,3,4$. Nevertheless, the semi-local terms contribute non-trivially to the correlation function in such a way that one recovers \eqref{e:TTTTfromV} exactly. Note also that \eqref{e:allvanish} is valid only in the parametrisation $g_{\mu\nu} = e^{-2\tau} \delta_{\mu\nu}$.

In order to check our results we have carried out all calculations in the parametrisation of the metric $g_{\mu\nu} = \Omega^2 \delta_{\mu\nu}$ as well. We present the computation in appendix \ref{sec:varphi}. All results including \eqref{e:TTT} and \eqref{e:TTTTfromV} are confirmed.

\section{Higher-point functions} \label{sec:higher}

We show in this short section that there is no scale violation in all connected higher point correlation functions of the trace of the stress-energy tensor. This is a consequence of the fact that anomalies are local and
the scaling dimension of the $n$-point function of the virial current equals $\Delta = (d-1)n - (n-1)d = d - n$ and becomes negative for $n > d$. Therefore,
\begin{equation}
\hdelta_\sigma \< V^{\mu_1} \ldots V^{\mu_n} \> = 0, \quad n \geq d
\end{equation}
and hence
\begin{equation}
\hdelta_\sigma \< \underbrace{T \ldots T}_{n} \> = 0, \quad n \geq d
\end{equation}
as well.

\section{Conclusions} \label{sec:con}

In this paper we analysed the structure of the scale anomaly in four dimensional unitary scale invariant theories. We found that
2-, 3-, and connected 4-point functions of the trace of the stress-energy tensor $T$ are anomalous while the anomaly in all connected higher point functions vanishes. The 2-point function of $T$ is non-trivial if and only if scale transformations are anomalous. It follows that
a unitary SFT is a CFT iff the scale anomaly vanishes\footnote{If the theory contains dimension 2 operators one may need to improve $T$ first.}(since then $T=0$ and this implies that the theory is conformal).

One of our main results is the explicit form of the anomaly in 3- and 4-point functions. The explicit expressions are given in \eqref{e:TTT} and \eqref{e:TTTTfromV} and were derived using the Wess-Zumino action and a careful treatment of semi-local terms
(terms with support on a set that contains both coincident and separated points). We also obtained the form of the anomaly both for 3- and 4-point functions by an independent computation using only Lorentz invariance, scale invariance and the fact that the anomaly is local.
This is presented in appendix \ref{sec:p2}

To obtain the semi-local contributions we computed all couplings of sources to operators that contribute up to 4-point functions. These terms are is given in \eqref{e:Sint2} or in alternative parametrisation in \eqref{e:Sintvp} and \eqref{c_coef}. The non-linear terms in sources encode the semi-local contributions to correlation functions. We emphasise that only after including all semi-local  contributions the final answer is independent of the parametrisation of the sources. For example, if one uses $\tau$ ($g_{\mu \nu} = e^{-2 \tau} \delta_{\mu \nu})$ as the source for $T$ then there is a contribution to the anomaly of the 4-point function that comes from the Wess-Zumino term, even in the on-shell forward scattering limit. On the other hand, if one uses $\varphi$ ($g_{\mu \nu} = (1+\varphi)^2 \delta_{\mu \nu}$) as the source for $T$, the contribution from
Wess-Zumino term in the on-shell forward scattering limit vanishes and there are additional contributions from semi-local terms, leading to the same answer.

In \cite{Luty} it was argued that the structure of the anomaly of the 3-point function is not compatible with OPEs and this then implies that the coefficient of the scale anomaly must vanish and thus all unitary SFTs are CFTs. We discussed here a subtlety in the relation between OPEs and the large momentum limit which invalidates this argument. While the OPE controls the leading non-local contribution in the large momentum limit, there are semi-local contributions which dominate over the OPE contribution in the relevant case. A detailed discussion of this subtlety is presented in appendix \ref{sec:large_momentum_limit}. Taken the semi-local terms into account one can no longer conclude that the scale anomaly coefficient must vanish.

In \cite{Dymarsky} it was argued that all dilaton amplitudes vanish in an on-shell forward scattering limit. As just reviewed, we find that
the scale anomaly of the 4-point function is non-zero in this limit. Nevertheless one cannot conclude (without additional assumptions) from the vanishing of the
amplitudes that the coefficient of the scale anomaly must vanish. One can only conclude that the 4-point function is semi-local in that limit. Of course, this by itself is a very strong constraint on the structure of the SFT.

All in all,  additional work is required in order to either  prove that four dimensional unitary SFTs and CFTs  or find a counterexample.

\bigskip

\noindent \textbf{Acknowledgments:} We would like to thank Yegor Korovin and especially Markus Luty for discussions. K.S.~acknowledges support from a grant of the John Templeton Foundation. A.B.~is supported by the Interuniversity Attraction Poles Programme initiated by the Belgian Science Policy (P7/37) and the Odysseus program of the FWO. The opinions expressed in this publication are those of the authors and do not necessarily reflect the views of the John Templeton Foundation.

\newpage
\appendix

\section{Large momentum limit and OPEs} \label{sec:large_momentum_limit}

In this section we argue that in general
\begin{equation} \label{e:op}
\lla \O_1(\bs{q}) \O_2(-\bs{q}+\bs{p}) \O_3(-\bs{p}) \rra \propto \left\{ \begin{array}{ll}
q^{\Delta_1 + \Delta_2 - \Delta_3 - d} p^{2 \Delta_3 - d} \left( 1 + o \left( p/q \right) \right) & \text{if } \Delta_3 < \frac{d}{2} \\
q^{\Delta_1 + \Delta_2 + \Delta_3 - 2 d} \left( 1 + o \left( p/q \right) \right) & \text{if } \Delta_3 > \frac{d}{2} \end{array} \right.
\end{equation}
in the limit $q \gg p$. As mentioned in the introduction, the behaviour presented in the first line of \eqref{e:op} will be called \emph{a naive OPE} behaviour, since it follows directly from the Fourier transform of the appropriate OPE term.

In the following subsections we analyse the limit $q \gg p$ in the context of CFTs. In the first subsection we present an example that demonstrates \eqref{e:op} for both cases. The example also illustrate that the semi-local terms cannot be removed by local counterterms (as expected) and thus cannot be ignored.  In subsection \ref{sec:fourier} we explain the difference between the two cases in
(\ref{e:op}) by direct Fourier transform of the 3-point function of three scalar operators and in subsection \ref{sec:3k}
we prove \eqref{e:op} using the triple-$K$ representation of CFT 3-point functions \cite{Bzowski:2013sza}.
These results show that \eqref{e:op} is valid in the conformal case and we expect that it would also hold in unitary SFTs, if such theories exist.

\subsection{Example}

Our aim in this subsection is to demonstrate \eqref{e:op}. The example is chosen such that the 3-point functions are given by simple expression in momentum space. Consider the correlation functions of scalar conformal primaries $\O_{3/2}$ and $\O_{5/2}$ of dimensions $\Delta = \frac{3}{2}$ and $\Delta = \frac{5}{2}$, respectively,  in a four dimensional CFT. The correlation functions are given by
\begin{align}
\lla \O_{3/2}(\bs{p}_1) \O_{3/2}(\bs{p}_2) \O_{3/2}(\bs{p}_3) \rra & = \frac{C_{3/2}}{p_1 p_2 p_3 \sqrt{p_1 + p_2 + p_3}}, \\
\lla \O_{5/2}(\bs{p}_1) \O_{5/2}(\bs{p}_2) \O_{5/2}(\bs{p}_3) \rra & = \frac{C_{5/2}}{\sqrt{p_1 + p_2 + p_3}},
\end{align}
where $C_{3/2}$ and $C_{5/2}$ are constants. These expression can be obtained by starting from the triple-$K$ representation of the correlators given in \cite{Bzowski:2013sza} and then carrying out the remaining integral\footnote{The triple-$K$ integral is elementary because for operators with half-integer dimension the corresponding $K$ Bessel functions reduce to elementary functions.}.

It follows that in the large momentum limit, the first correlation function yields
\begin{equation}
\lla \O_{3/2}(\bs{q}) \O_{3/2}(-\bs{q}+\bs{p}) \O_{3/2}(-\bs{p}) \rra = \frac{C_{3/2}}{\sqrt{2} q^{\frac{5}{2}} p} + \ldots,
\end{equation}
consistent with the first line of \eqref{e:op}, since $\frac{3}{2} < \frac{d}{2}$, while
\begin{equation} \label{o5/2}
\lla \O_{5/2}(\bs{q}) \O_{5/2}(-\bs{q}+\bs{p}) \O_{5/2}(-\bs{p}) \rra = \frac{C_{5/2}}{\sqrt{2 q}} + \ldots
\end{equation}
which is consistent with the second line of \eqref{e:op}, since in this case $\frac{5}{2} > \frac{d}{2}$. Note, however, that the leading term is local in the sense that up to a constant
\begin{equation} \label{semi}
\int \D^4 \bs{x}_1 \D^4 \bs{x}_2 \D^4 \bs{x}_3 e^{- \I \bs{p}_1 \cdot \bs{x}_1} e^{- \I \bs{p}_2 \cdot \bs{x}_2} e^{- \I \bs{p}_3 \cdot \bs{x}_3} \frac{1}{|\bs{x}_1 - \bs{x}_3|^{\frac{7}{2}}} \delta(\bs{x}_1 - \bs{x}_2) \propto \delta(\bs{p}_1 + \bs{p}_2 + \bs{p}_3) \frac{1}{\sqrt{p_3}}.
\end{equation}
Nevertheless, it cannot be removed by any local counterterm\footnote{Actually in this case there are no local terms of dimension 4 that one can construct using the source $\phi_0$ associated to the operator $\O_{5/2}$ since $\phi_0$ has dimension 3/2.}. This is so because
this term has support at $\bs{x}_1=\bs{x}_2 \neq \bs{x}_3$ (terms with support at $\bs{x}_1 =\bs{x}_2=\bs{x}_3$ are analytic in all momenta and (\ref{semi}) is non-analytic in $p_3$)  and the contribution from any local counterterms would have support at $\bs{x}_1 =\bs{x}_2=\bs{x}_3$. Only ultra-local terms can be removed by local counterterms and this is an example of a semi-local term.

\subsection{The Fourier transform} \label{sec:fourier}

In this subsection we show how \eqref{e:op} emerges by taking a direct Fourier transform of the 3-point function of scalar conformal primaries.

Let us consider three scalar conformal primaries $\O_1, \O_2, \O_3$ of dimensions $\Delta_1, \Delta_2, \Delta_3$ respectively. The exact 3-point function in position space reads
\begin{equation} \label{e:3pt}
\< \O_1(\bs{x}_1) \O_2(\bs{x}_2) \O_3(\bs{x}_3) \> = \frac{C_{123} C_{33}}{|\bs{x}_1 - \bs{x}_2|^{\Delta_1 + \Delta_2 - \Delta_3} |\bs{x}_2 - \bs{x}_3|^{\Delta_2 + \Delta_3 - \Delta_1} |\bs{x}_3 - \bs{x}_1|^{\Delta_3 + \Delta_1 - \Delta_2}},
\end{equation}
where $C_{33}$ is a normalisation of the 2-point function
\begin{equation}
\< \O_3(\bs{x}_1) \O_3(\bs{x}_2) \> = \frac{C_{33}}{|\bs{x}_1 - \bs{x}_2|^{2 \Delta_3}}
\end{equation}
and $C_{123}$ is the OPE coefficient
\begin{equation} \label{e:1}
\O_1(\bs{x}_1) \O_2(\bs{x}_2) \sim \frac{C_{123}}{|\bs{x}_1 - \bs{x}_2|^{\Delta_1 + \Delta_2 - \Delta_3}} \O_3(\bs{x}_2) + \ldots
\end{equation}
Now one can carry out the Fourier transform of \eqref{e:3pt} in the large momentum limit $p_1, p_2 \gg p_3$. As we will see, it does not invalidate the statement that the leading momentum behaviour comes from the region where $\bs{x}_1$ is close to $\bs{x}_2$: instead it shows that an additional contribution to the singularity at $\bs{x}_1 = \bs{x}_2$ may appear when the Fourier transform over $\bs{x}_3$ is carried out.

The Fourier transform is given by
\begin{align} \label{e:2}
\< \O_1(\bs{p}_1) \O_2(\bs{p}_2) \O_3(\bs{p}_3) \> & = \int \D^d \bs{x}_1 \D^d \bs{x}_2 \D^d \bs{x}_3  \frac{C_{123} C_{33} e^{-\I \bs{p}_1 \cdot \bs{x}_1} e^{-\I \bs{p}_2 \cdot \bs{x}_2} e^{-\I \bs{p}_3 \cdot \bs{x}_3}}{|\bs{x}_1 - \bs{x}_2|^{\Delta_1 + \Delta_2 - \Delta_3} |\bs{x}_2 - \bs{x}_3|^{\Delta_2 + \Delta_3 - \Delta_1} |\bs{x}_3 - \bs{x}_1|^{\Delta_3 + \Delta_1 - \Delta_2}} \nonumber\\
& = \int \D^d \bs{x}_1 \D^d \bs{x}_2  \frac{C_{123} C_{33} e^{-\I \bs{p}_1 \cdot \bs{x}_1} e^{-\I \bs{p}_2 \cdot \bs{x}_2}}{|\bs{x}_1 - \bs{x}_2|^{\Delta_1 + \Delta_2 - \Delta_3}} F(\bs{x}_1 - \bs{x}_2, \bs{p}_3),
\end{align}
where in the second line the integral over $\bs{x}_3$ was carried out,
\begin{equation} \label{e:3}
F(\bs{x}_1 - \bs{x}_2, \bs{p}_3) = \int \D^d \bs{x}_3 \frac{e^{-\I \bs{p}_3 \cdot \bs{x}_3}}{|\bs{x}_2 - \bs{x}_3|^{\Delta_2 + \Delta_3 - \Delta_1} |\bs{x}_3 - \bs{x}_1|^{\Delta_3 + \Delta_1 - \Delta_2}}.
\end{equation}
Note that the factor $|\bs{x}_1 - \bs{x}_2|^{-(\Delta_1 + \Delta_2 - \Delta_3)}$ in \eqref{e:2} is exactly equal to the factor in OPE \eqref{e:1}. Therefore, the naive OPE is valid if $F$ is regular at $\bs{x}_1 = \bs{x}_2$. Otherwise, $F$ contributes an additional singularity to the integrals over $\bs{x}_1$ and $\bs{x}_2$.

We can find out what is the leading behaviour of $F$ with respect to $|\bs{x}_1 - \bs{x}_2|$ and $\bs{p}_3$. Taking $\bs{x}_1 = \bs{x}_2$ in \eqref{e:3} we find
\begin{equation} \label{e:4}
F(0, \bs{p}_3) = \int \D^d \bs{x}_3 \frac{e^{-\I \bs{p}_3 \cdot \bs{x}_3}}{x_3^{2 \Delta_3}} \propto p_3^{2 \Delta_3 - d},
\end{equation}
which converges if $2 \Delta_3 < d$. Therefore, we have shown that the naive OPE expansion is valid only if $\Delta_3 < \frac{d}{2}$,
\begin{equation}
\lla \O_1(\bs{q}) \O_2(-\bs{q}+\bs{p}) \O_3(-\bs{p}) \rra \propto p^{2 \Delta_3 - d} q^{\Delta_1 + \Delta_2 - \Delta_3 - d} \left( 1 + o \left(p/q \right) \right).
\end{equation}
On the other hand if $2 \Delta_3 \geq d$, then $F(0, \bs{p}_3) = \infty$, or in other words $F(\bs{x}_1 - \bs{x}_2, \bs{p}_3)$ is singular at $\bs{x}_1 = \bs{x}_2$. This means that the integral over $\bs{x}_3$ contributes an additional singularity at $\bs{x}_1 = \bs{x}_2$ and this modifies the large $q$ limit. To obtain the answer in this case we need to start from the 3-point function computed directly in momentum space and we do this in the next subsection.

\subsection{\texorpdfstring{Proof of \eqref{e:op}}{Proof of (A.1)}} \label{sec:3k}

We can obtain the large momentum limit \eqref{e:op} in all cases by starting from the momentum space representation of the correlators derived in \cite{Bzowski:2013sza}. The 3-point functions of scalar operators in any CFT can be represented by the \emph{triple-$K$} integral
\begin{align}
& \lla \mathcal{O}_1(\bs{p}_1) \mathcal{O}_2(\bs{p}_2) \mathcal{O}_3(\bs{p}_3)
\rra \nn\\
& \qquad = C_{123}  p_1^{\Delta_1 - \frac{d}{2}} p_2^{\Delta_2 - \frac{d}{2}}
p_3^{\Delta_3 - \frac{d}{2}} \int_0^\infty \D x \: x^{\frac{d}{2} - 1}
K_{\Delta_1 - \frac{d}{2}} (p_1 x) K_{\Delta_2 - \frac{d}{2}} (p_2 x)
K_{\Delta_3 - \frac{d}{2}} (p_3 x), \label{scalar_3pt}
\end{align}
where $K_\nu(p)$ is a modified Bessel function of the second kind (or Bessel $K$ function, for short) and $C_{123}$ is an overall undetermined constant. Using such a representation it is easy to consider the large momentum limit $q = p_1 \cong p_2 \gg p_3 = p$. To do it, fix the value of $q$ and expand the integrand in \eqref{scalar_3pt} as a power series in $p$, according to
\begin{equation} \label{e:expK}
p^{\nu} K_{\nu}(p x) = \left[ \Gamma (\nu) 2^{\nu-1} x^{-\nu} + O(p^2) \right] + \left[ p^{2 \nu} \Gamma( - \nu ) 2^{-\nu - 1} x^\nu + O(p^{2 \nu + 2}) \right].
\end{equation}
Since $\nu = \Delta_3 - \frac{d}{2}$, we can see that the form of the leading term in $p/q$ depends on whether $2 \Delta_3 < d$ or $2 \Delta_3 > d$. One can combine the two cases by writing
\begin{equation} \label{e:expK1}
p^{\nu} K_{\nu}(p x) = \Gamma ( | \nu | ) 2^{|\nu| - 1} x^{-|\nu|} p^{2 \nu \theta(-\nu)} + \ldots,
\end{equation}
where $\theta$ denotes the step function: $\theta(x) = 1$ for $x \geq 0$ and $\theta(x) = 0$ for $x < 0$. The remaining integral can be evaluated explicitly by means of the formula
\begin{align} \label{e:I2K}
& \int_0^\infty \D x \: x^{\alpha - 1} K_\mu(q x) K_\nu(q x) = \nn\\
& \qquad = \frac{2^{\alpha - 3}}{\Gamma(\alpha) q^{\alpha}} \Gamma \left( \frac{\alpha + \mu + \nu}{2} \right) \Gamma \left( \frac{\alpha + \mu - \nu}{2} \right) \Gamma \left( \frac{\alpha - \mu + \nu}{2} \right) \Gamma \left( \frac{\alpha - \mu - \nu}{2} \right),
\end{align}
see \cite{Bzowski:2013sza} for details. The dichotomy in the expansion \eqref{e:expK} is the primary reason for the occurrence of the two cases in \eqref{e:op}. By substituting \eqref{e:expK1} into the triple-$K$ integral \eqref{scalar_3pt} and using \eqref{e:I2K} one finds
\begin{align}
& \lim_{q \gg p} \lla \mathcal{O}_1(\bs{q}) \mathcal{O}_2(-\bs{q}+\bs{p}) \mathcal{O}_3(-\bs{p})
\rra = \nn\\
& \qquad = C_{123} C_0(\Delta_i, d) q^{\Delta_1 + \Delta_2 + \left| \Delta_3 - \frac{d}{2} \right| - \frac{3 d}{2}} p^{(2 \Delta_3 - d) \theta \left( \frac{d}{2} - \Delta_3 \right)} \left( 1 + o \left(p/q \right) \right),
\end{align}
where $C_0(\Delta_i, d)$ is a specific numerical constant. This expression coincides with \eqref{e:op} when the step function and the absolute value are resolved into particular cases.

Assume now $\nu = \Delta_3 - \frac{d}{2} > 0$. As pointed out in section \ref{sec:intro2}, the leading term in the large momentum expansion is in this case semi-local, \textit{i.e.}, it is a Fourier transform of a position space expression supported on the set of coincident points, for example
\begin{equation}
\delta(\bs{x}_1 - \bs{x}_3) \frac{1}{| \bs{x}_1 - \bs{x}_2|^{\Delta_1 + \Delta_2 + \Delta_3 - d}} \quad \stackrel{\mathcal{F}}{\longmapsto} \quad q^{\Delta_1 + \Delta_2 + \Delta_3 - 2 d} \left( 1 + o \left(p/q \right) \right).
\end{equation}
Using the triple-$K$ integral representation one can show that all subleading terms up to, but excluding, $q^{\Delta_1 + \Delta_2 - \Delta_3 - d} p^{2 \Delta_3 - d}$ are semi-local in the large momentum limit. This follows from the Taylor expansion \eqref{e:expK}, which in general takes form
\begin{equation}
p^{\nu} K_{\nu}(p x) = x^{-\nu} \sum_{j=0}^\infty a_j (p x)^{2j} + p^{2 \nu} x^\nu \sum_{j=0}^\infty b_j (p x)^{2j},
\end{equation}
where the series coefficients $a_j$ and $b_j$ are known, see \textit{e.g.}, \cite{Abramowitz}. By substituting this result to \eqref{scalar_3pt} and using \eqref{e:I2K} one finds the expansion of the triple-$K$ integral in the large momentum limit,
\begin{align} \label{e:expansion}
& \lim_{q \gg p} \lla \mathcal{O}_1(\bs{q}) \mathcal{O}_2(-\bs{q}+\bs{p}) \mathcal{O}_3(-\bs{p}) \rra = C_{123} \left[ \sum_{j=0}^{\infty} C_j(\Delta_i, d) q^{\Delta_1 + \Delta_2 + \Delta_3 - 2d - 2j} p^{2j} \right.\nn\\
& \qquad\qquad \left. + \: \sum_{j=0}^{\infty} D_j(\Delta_i, d) q^{\Delta_1 + \Delta_2 - \Delta_3 - d - 2j} p^{2 \Delta_3 - d + 2j} \right],
\end{align}
where $C_j$ and $D_j$ represent some numerical constants explicitly computable by means of \eqref{e:I2K}. As one can see, the first series contains only even powers of momentum $p$. In position space all such terms are semi-local, \textit{i.e.}, they are Fourier transforms of the form
\begin{equation}
\Box_{\bs{x}_3}^j \delta(\bs{x}_1 - \bs{x}_3) \frac{1}{| \bs{x}_1 - \bs{x}_2|^{\Delta_1 + \Delta_2 + \Delta_3 - d - 2j}} \quad \stackrel{\mathcal{F}}{\longmapsto} \quad q^{\Delta_1 + \Delta_2 + \Delta_3 - 2 d - 2j} p^{2j} \left( 1 + o \left(p/q \right) \right)
\end{equation}
in the large momentum limit, up to a multiplicative constant. As long as $j < 2 \nu = 2 \Delta_3 - d$ these terms are more leading  than the terms in the second series in the expansion \eqref{e:expansion}. As one can see the terms featuring in the second power series are not semi-local and the leading $D_0$ term reproduces the naive OPE term in \eqref{e:op}.

\section{Alternative derivation of the anomaly} \label{sec:p2}

In this appendix we show how to obtain the form of the scale violation \eqref{e:TTT} and \eqref{e:TTTTfromV} directly from Lorentz and scale invariance (plus the locality of anomalies) applied to the 3- and 4-point functions of the virial current.

Let us start with the 3-point function.  Following \cite{Bzowski:2013sza} we find that the most general tensor decomposition of $\< V^{\mu_1} V^{\mu_2} V^{\mu_3} \>$ is given by
\begin{align} \label{e:predecomp}
& \I \lla V^{\mu_1}(\bs{p}_1) V^{\mu_2}(\bs{p}_2) V^{\mu_3}(\bs{p}_3) \rra = \pi^{\mu_1}_{\alpha_1}(\bs{p}_1) \pi^{\mu_2}_{\alpha_2}(\bs{p}_2) \pi^{\mu_3}_{\alpha_3}(\bs{p}_3) T_1^{\alpha_1 \alpha_2 \alpha_3} \nn\\
& \qquad + \: \frac{p^{\mu_1}_1}{p_1^2} \pi^{\mu_2}_{\alpha_2}(\bs{p}_2) \pi^{\mu_3}_{\alpha_3}(\bs{p}_3) T_2^{\alpha_2 \alpha_3} + 2 \text{ permutations} \nn\\
& \qquad + \: \frac{p^{\mu_2}_2}{p_2^2} \frac{p^{\mu_3}_3}{p_3^2} \pi^{\mu_1}_{\alpha_1}(\bs{p}_1) T_3^{\alpha_1} + 2 \text{ permutations} \nn\\
& \qquad + \: \frac{p^{\mu_1}_1}{p_1^2} \frac{p^{\mu_2}_2}{p_2^2} \frac{p^{\mu_3}_3}{p_3^2} T_4.
\end{align}
where $\pi^\mu_\alpha(\bs{p}) = \delta^\mu_\alpha - p^\mu p_\alpha/p^2$ is a transverse projector and
the tensors $T_1$ through $T_4$ are built using the metric $\delta_{\mu \nu}$ and the momenta.
Due to momentum conservation $\bs{p}_1 + \bs{p}_2 + \bs{p}_3 = 0$ only two of the momenta are independent but we may choose different momenta for different indices. The following choice is convenient \cite{Bzowski:2013sza},
\begin{equation} \label{e:rule}
\bs{p}_1, \bs{p}_2 \text{ for } \alpha_1, \qquad \bs{p}_2, \bs{p}_3 \text{ for } \alpha_2, \qquad \bs{p}_3, \bs{p}_1 \text{ for } \alpha_3.
\end{equation}
With this rule all possible tensor forms in, for example, $T_2^{\alpha_2 \alpha_3}$ are
\begin{equation}
\delta^{\alpha_2 \alpha_3}, \qquad p_3^{\alpha_2} p_1^{\alpha_3}, \qquad p_3^{\alpha_2} p_3^{\alpha_3}, \qquad p_2^{\alpha_2} p_1^{\alpha_3}, \qquad p_2^{\alpha_2} p_3^{\alpha_3}.
\end{equation}
However, since $p^\alpha \pi^{\mu}_{\alpha}(\bs{p}) = 0$, the contraction of the last three tensors with the prefactor in \eqref{e:predecomp} vanishes. Hence only the first two tensors listed above may appear in $T_2^{\alpha_2 \alpha_3}$.

By carrying out the analysis for the remaining factors we find the most general parity even decomposition of the 3-point function to be
\begin{align} \label{e:decomp}
& \I \lla V^{\mu_1}(\bs{p}_1) V^{\mu_2}(\bs{p}_2) V^{\mu_3}(\bs{p}_3) \rra = \pi^{\mu_1}_{\alpha_1}(\bs{p}_1) \pi^{\mu_2}_{\alpha_2}(\bs{p}_2) \pi^{\mu_3}_{\alpha_3}(\bs{p}_3) \left[ A_1 p_2^{\alpha_1} p_3^{\alpha_2} p_1^{\alpha_3} \right.\nonumber\\
& \qquad\qquad\qquad \left. + \left( A_{21} p_2^{\alpha_1} \delta^{\alpha_2 \alpha_3} + A_{22} p_3^{\alpha_2} \delta^{\alpha_1 \alpha_3} + A_{23} p_1^{\alpha_3} \delta^{\alpha_1 \alpha_2} \right) \right] \nonumber\\
& \qquad + \: \frac{p^{\mu_1}_1}{p_1^2} \pi^{\mu_2}_{\alpha_2}(\bs{p}_2) \pi^{\mu_3}_{\alpha_3}(\bs{p}_3) \left( B_{11} p_3^{\alpha_2} p_1^{\alpha_3} + B_{21} \delta^{\alpha_2 \alpha_3} \right)
+ \frac{p^{\mu_2}_2}{p_2^2} \pi^{\mu_1}_{\alpha_1}(\bs{p}_1) \pi^{\mu_3}_{\alpha_3}(\bs{p}_3) \left( B_{12} p_2^{\alpha_1} p_1^{\alpha_3} + B_{22} \delta^{\alpha_1 \alpha_3} \right) \nn\\
& \qquad\qquad\qquad\qquad + \: \frac{p^{\mu_3}_3}{p_3^2} \pi^{\mu_1}_{\alpha_1}(\bs{p}_1) \pi^{\mu_2}_{\alpha_2}(\bs{p}_2) \left( B_{13} p_2^{\alpha_1} p_3^{\alpha_2} + B_{23} \delta^{\alpha_1 \alpha_2} \right) \nn\\
& \qquad + \: \frac{p^{\mu_2}_2}{p_2^2} \frac{p^{\mu_3}_3}{p_3^2} \pi^{\mu_1}_{\alpha_1}(\bs{p}_1) C_{1} p_2^{\alpha_1} + \frac{p^{\mu_1}_1}{p_1^2} \frac{p^{\mu_3}_3}{p_3^2} \pi^{\mu_2}_{\alpha_2}(\bs{p}_2) C_{2} p_3^{\alpha_2} + \frac{p^{\mu_1}_1}{p_1^2} \frac{p^{\mu_2}_2}{p_2^2} \pi^{\mu_3}_{\alpha_3}(\bs{p}_3) C_{3} p_1^{\alpha_3} \nn\\
& \qquad + \: \frac{p^{\mu_1}_1}{p_1^2} \frac{p^{\mu_2}_2}{p_2^2} \frac{p^{\mu_3}_3}{p_3^2} D,
\end{align}
where the form factors, $A_1, A_{2j}, B_{ij}, C_j, D\ (j{=}1,2,3, i{=}1,2)$,  are scalar functions of the momenta magnitudes $p_j = | \bs{p}_j |$. With this decomposition it follows that $\lla T(\bs{p}_1) T(\bs{p}_2) T(\bs{p}_3) \rra = D$ due to the relation $T = - \partial_\mu V^\mu$. Furthermore, if the virial current is of the form \eqref{e:improve} then the 3-point function should be purely longitudinal, {\it i.e.} all form factors other than $D$ must vanish or be at most local. Therefore, the problem of scale vs conformal invariance can be restated as the question of whether it is possible to have a non-conserved current of dimension $\Delta = 3$ in a scale invariant theory which has at least one non-local form factor among $A, B$ and $C$.

Each form factor in \eqref{e:decomp} has a specified scaling dimension, up to anomaly. Since the dimension of the entire correlation function equals $\Delta_{\text{tot}} = 3 \Delta - 2 d = 1$, one finds the following scaling dimensions for the form factors in momentum space,
\begin{equation} \label{e:dims}
\Delta(A_1) = -2, \qquad \Delta(A_{2j}) = \Delta(B_{1j}) = 0, \qquad \Delta(B_{2j}) = \Delta(C_j) = 2, \qquad \Delta(D) = 4.
\end{equation}
The form factors are not independent as they carry a representation of the permutation group. The symmetry properties follow from the symmetry of the 3-point function,
\begin{equation}
\lla V^{\mu_1}(\bs{p}_1) V^{\mu_2}(\bs{p}_2) V^{\mu_3}(\bs{p}_3) \rra = \lla V^{\mu_{\sigma(1)}}(\bs{p}_{\sigma(1)}) V^{\mu_{\sigma(2)}}(\bs{p}_{\sigma(2)}) V^{\mu_{\sigma(3)}}(\bs{p}_{\sigma(3)}) \rra
\end{equation}
for any permutation $\sigma$ of the set $\{1,2,3\}$. The action of the permutation $\sigma$ on a given form factor $F$ is
\begin{equation}
F^{(\sigma)} = F(p_{\sigma(1)}, p_{\sigma(2)}, p_{\sigma(3)}).
\end{equation}
By applying the symmetries to the decomposition \eqref{e:decomp} and requiring the invariance of the entire correlation function, we find
\begin{align}
A_1 & = 0, && D = D^{(\sigma)}, \nn\\
A_{2j} & = (-1)^\sigma A^{(\sigma)}_{2\sigma(j)}, && C_{j} = (-1)^\sigma C^{(\sigma)}_{\sigma(j)}, \nn\\
B_{nj} & = B^{(\sigma)}_{n\sigma(j)}, \label{e:symprop}
\end{align}
where $(-1)^\sigma$ denotes the sign of the permutation $\sigma$. The vanishing of the form factor $A_1$ is related to the well-known fact \cite{Osborn:1993cr} that a 3-point function of any Abelian conserved current in a CFT vanishes. Here, however, the current is not conserved and the theory is only assumed to be scale invariant. Hence, the remaining parts of the correlation functions can be non-vanishing.

The decomposition \eqref{e:decomp} can be expanded into a basis of simple tensors such as $p_2^{\mu_1} p_3^{\mu_2} p_1^{\mu_3}$, $\delta^{\mu_1 \mu_2} p_1^{\mu_3}$, \textit{etc}. As discussed before, we may choose two out of three independent momenta to appear under each Lorentz index. We stick to the rule \eqref{e:rule}, now applied to Lorentz indices $\mu_j$ instead of $\alpha_j$, $j=1,2,3$. In this case it is relatively easy to connect the form factors appearing in \eqref{e:decomp} to the coefficients of simple tensors. In particular we find
\begin{align} \label{e:decomp2}
& \I \lla V^{\mu_1}(\bs{p}_1) V^{\mu_2}(\bs{p}_2) V^{\mu_3}(\bs{p}_3) \rra = 0 \times p_2^{\mu_1} p_3^{\mu_2} p_1^{\mu_3} + A_{23} \delta^{\mu_1 \mu_2} p_1^{\mu_3} \nn\\
& \qquad + \: \frac{\delta^{\mu_1 \mu_2} p_3^{\mu_3}}{p_3^2} \left[ B_{23} + \tfrac{1}{2} A_{23} ( p_1^2 - p_2^2 + p_3^2 ) \right] \nn\\
& \qquad + \: \frac{p_2^{\mu_1} p_3^{\mu_2} p_3^{\mu_3}}{p_3^2} \left[ B_{13} + A_{22} - A_{21} \right]  \nn\\
& \qquad + \: \frac{p_2^{\mu_1} p_2^{\mu_2} p_3^{\mu_3}}{2 p_2^2 p_3^2} \left[ 2 ( C_1 + B_{22} - B_{23} ) + (B_{13} + A_{22}) (p_2^2 + p_3^2 - p_1^2) \right.\nn\\
& \qquad\qquad\qquad \left. + \: (B_{12} + A_{21} - A_{23}) (p_1^2 - p_2^2 + p_3^2) \right]  \nn\\
& \qquad + \: \frac{p_1^{\mu_1} p_2^{\mu_2} p_3^{\mu_3}}{4 p_1^2 p_2^2 p_3^2} \left[ 4 D + 2 (C_1 + B_{22}) (p_1^2 + p_2^2 - p_3^2) \right. \nn\\
& \qquad\qquad\qquad + \: 2 (C_2 + B_{23}) (p_2^2 + p_3^2 - p_1^2) + 2 (C_3 + B_{21}) (p_1^2 + p_3^2 - p_2^2) \nn\\
& \qquad\qquad\qquad + \: (A_{21} + B_{12}) (p_1^2 + p_2^2 - p_3^2) (p_1^2 - p_2^2 + p_3^2) \nn\\
& \qquad\qquad\qquad + \: (A_{22} + B_{13}) (- p_1^2 + p_2^2 + p_3^2) (p_1^2 + p_2^2 - p_3^2) \nn\\
& \qquad\qquad\qquad \left. + \: (A_{23} + B_{11}) (- p_1^2 + p_2^2 + p_3^2) (p_1^2 - p_2^2 + p_3^2) \right] \nn\\
& \qquad + \: \ldots
\end{align}
where the omitted terms do not contain the tensors listed explicitly.
The only terms that may contain scale violating expressions are terms of non-negative scaling dimension, as follows from the locality of anomalies. Since the total dimension of the correlation function is $\Delta_{\text{tot}} = 1$, such terms cannot appear in front of a tensor containing three momenta. Therefore, while the coefficients $B_{13}, A_{21}$ and $A_{22}$ may contain logarithms, the combined coefficient of $p_2^{\mu_1} p_3^{\mu_2} p_3^{\mu_3}$ cannot. Hence the third line of \eqref{e:decomp2} requires that $\hdelta_\sigma B_{13} = \hdelta_\sigma A_{21} - \hdelta_\sigma A_{22}$. One can substitute this result back to \eqref{e:decomp2} and read off the equation following from the requirement that the coefficient of $p_2^{\mu_1} p_2^{\mu_2} p_3^{\mu_3}$ is scale invariant. Using the symmetry properties \eqref{e:symprop} one finds
\begin{equation}
\hdelta_\sigma C_1 = \hdelta_\sigma B_{23} - \hdelta_\sigma B_{22} - \frac{1}{2} (p_2^2 + p_3^2 - p_1^2) \hdelta_\sigma A_{21}.
\end{equation}
Finally, this result together with the requirement that the coefficient of $p_1^{\mu_1} p_2^{\mu_2} p_3^{\mu_3}$ in \eqref{e:decomp2} is scale invariant leads to
\begin{equation} \label{e:Dviol}
\hdelta_\sigma D = \frac{1}{2} \left[ ( p_1^2 - p_2^2 - p_3^2) \hdelta_\sigma B_{21} + ( p_2^2 - p_1^2 - p_3^2 ) \hdelta_\sigma B_{22} + ( p_3^2 - p_1^2 - p_2^2 ) \hdelta_\sigma B_{23} \right].
\end{equation}
We can further constrain the form of scale violating part of the form factors $B_{2j}$ by looking at the symmetry properties \eqref{e:symprop}. Specifically, consider the form factor $B_{23}$, which is antisymmetric under the exchange $p_1 \leftrightarrow p_2$. From \eqref{e:dims} one sees that its scaling dimension equals two and hence its most general scale violation is
\begin{equation} \label{dB}
\hdelta_\sigma B_{23} = c p_3^2 + c_1 ( p_2^2 - p_1^2 ),
\end{equation}
where $c$ and $c_1$ are two undetermined constants. The scale violations $\hdelta_\sigma B_{21}$ and $\hdelta_\sigma B_{22}$ follow from \eqref{dB} using (\ref{e:symprop}).
By substituting back to \eqref{e:Dviol} one finds that the terms with $c_1$ cancel out and
the most general form of the scale violation in the 3-point function of the trace of the stress-energy tensor is,
\begin{equation} \label{e:anomal2}
\hdelta_\sigma \lla T(\bs{p}_1) T(\bs{p}_2) T(\bs{p}_3) \rra = \hdelta_\sigma D = - \tfrac{1}{2} c \sigma J^2,
\end{equation}
where $J^2$ is given by \eqref{e:J2}. Our analysis shows that the scale violating terms in the 3-point function of $\< T T T \>$ in any scale invariant theory in $d=4$ is constrained to take the form \eqref{e:TTT}. The value of the undetermined coefficient $c$ ({\it i.e.} that it is equal to -$4 e_{TT}$) cannot be determined without further input, such as the Wess-Zumino action.

A similar method can be applied to the 4-point function of the virial current.
In this case, however, the computation is much simpler. This is due to the fact that the total dimension of $\lla V^{\mu_1} V^{\mu_2} V^{\mu_3} V^{\mu_4} \rra$ in momentum space equals zero and hence the only scale violating form factors are those multiplying simple tensors containing metrics only. Due to the symmetry of the correlation function, there exists a unique tensor \eqref{e:Stensor} with such properties and the scale violation must take the form
\begin{equation}
\hdelta_\sigma \lla V^{\mu_1}(\bs{p}_1) V^{\mu_2}(\bs{p}_2) V^{\mu_3}(\bs{p}_3) V^{\mu_4}(\bs{p}_4) \rra = c \sigma \left( \delta^{\mu_1 \mu_2} \delta^{\mu_3 \mu_4} + \delta^{\mu_1 \mu_3} \delta^{\mu_2 \mu_4} + \delta^{\mu_1 \mu_4} \delta^{\mu_2 \mu_3} \right),
\end{equation}
for some numerical constant $c$. Therefore, based on Lorentz and scale invariance as well as the locality of anomalies one can deduce that the most general form of the scale violation in the 4-point function of the trace of the stress-energy tensor reads
\begin{align}
& \hdelta_\sigma \lla T(\bs{p}_1) T(\bs{p}_2) T(\bs{p}_3) T(\bs{p}_4) \rra = c \sigma  \times \nn\\
& \qquad \times \left[ (\bs{p}_1 \cdot \bs{p}_2) (\bs{p}_3 \cdot \bs{p}_4) + (\bs{p}_1 \cdot \bs{p}_3) (\bs{p}_2 \cdot \bs{p}_4) + (\bs{p}_1 \cdot \bs{p}_4) (\bs{p}_2 \cdot \bs{p}_3) \right].
\end{align}
This form of the 4-point functions was confirmed by direct calculations in sections \ref{sec:VVVV} and \ref{sec:TTTT}.  The value of the undetermined coefficient $c$ ({\it i.e.} that it is equal to -$8 ( e_{TT} + c_2^2 e_{22}/4)$) cannot be determined without further input, such as the Wess-Zumino action.

\section{\texorpdfstring{Calculations with $g_{\mu\nu}=(1 + \varphi)^2 \delta_{\mu\nu}$}{Calculations with g[ij] = (1 + f)**2 delta[ij]}} \label{sec:varphi}

In this section we outline the result of calculations of the 3- and 4-point functions in the representation of the metric
\begin{equation}
g_{\mu\nu} = \Omega^2 \delta_{\mu\nu}, \qquad\qquad \Omega = 1 + \varphi
\end{equation}
In this representation the Ricci scalar reads
\begin{equation}
R[\Omega^2 \delta_{\mu\nu}] = - (d - 1) \left[ 2 \Omega^{-3} \partial^2 \Omega + (d - 4) \Omega^{-4} (\partial \Omega)^2 \right]
\end{equation}
and for the trace of the stress-energy tensor we find
\begin{equation}
\Omega \frac{\delta}{\delta \Omega} = - 2 g^{\mu\nu} \frac{\delta}{\delta g^{\mu\nu}}, \qquad \frac{\delta}{\delta \Omega} = \frac{\delta}{\delta \varphi}, \qquad T = \frac{-1}{\Omega^{d-1}} \frac{\delta S}{\delta \Omega}.
\end{equation}
Furthermore, the interaction action \eqref{e:Sint} may be parametrised as
\begin{align} \label{e:Sintvp}
S_{\text{int}} & = \int \D^4 \bs{x} \left[ - \varphi T + C_{\mu} V^{\mu} + \phi_2 \O_2 + \phi_0 \O_4 + \ldots \right.\nn\\
& \qquad + \: \tfrac{1}{2} \varphi^2 \left( c^{\vp}_T T + c'^{\vp}_2 \partial^2 \O_2 + c^{\vp}_4 \O_4 \right) + \tfrac{1}{2} c^{\vp}_2 (\partial \varphi)^2 \O_2 \nn\\
& \qquad \left. + \: \tfrac{1}{2} \tilde{c}_{2}^{\vp} C_\mu C^\mu \O_2 + \tfrac{1}{6} \varphi^3 c^{(3)\vp}_T + \ldots \right].
\end{align}
By expanding $\varphi = - \tau + \tfrac{1}{2} \tau^2 - \tfrac{1}{6} \tau^3 + O(\tau^4)$, one finds
\begin{equation} \label{c_coef}
c_2^{\vp} = \tilde{c}_2^{\vp} = c_2, \qquad c_4^{\vp} = c_2'^{\vp} = 0, \qquad c_T^{\vp} = 1, \qquad c_T^{(3)\vp} = -2.
\end{equation}
These results can be recovered by independent calculations following the same lines as in the main text. Furthermore, since the reparametrisation of the metric does not alter other couplings, the result for correlation functions involving the virial current remains unchanged.

\subsection{\texorpdfstring{$\< TTT \>$}{< TTT >}}

We follow the same steps as in section \ref{sec:TTT}. In the new parametrisation
\begin{equation}
\frac{\delta T(\bs{x}_1)}{\delta \varphi(\bs{x}_2)} = - (d-1) \delta(\bs{x}_1 - \bs{x}_2) T - \frac{\delta^2 S_{\text{int}}}{\delta \varphi(\bs{x}_1) \delta \varphi(\bs{x}_2)}
\end{equation}
and hence the counterpart of \eqref{e:TTT3} reads
\begin{align}
& \< T(\bs{x}_1) T(\bs{x}_2) T(\bs{x}_3) \> = \frac{\delta^3 W}{\delta \vp(\bs{x}_1) \delta \vp(\bs{x}_2) \delta \vp(\bs{x}_3)} \nn\\
& \qquad + \: c^{\vp}_T \left[ \delta(\bs{x}_1 - \bs{x}_2) \< T(\bs{x}_2) T(\bs{x}_3) \> + \delta(\bs{x}_2 - \bs{x}_3) \< T(\bs{x}_3) T(\bs{x}_1) \> + \delta(\bs{x}_3 - \bs{x}_1) \< T(\bs{x}_1) T(\bs{x}_2) \> \right].
\end{align}
The scale violation from the Wess-Zumino action, however, is different in the new parametrisation of the metric. In total one finds,
\begin{align}
& \hdelta_\sigma \lla T(\bs{p}_1) T(\bs{p}_2) T(\bs{p}_3) \rra =  2 \sigma e_{TT} J^2 - 2 \sigma (c^{\vp}_T - 1) e_{TT} (p_1^4 + p_2^4 + p_3^4),
\end{align}
where $J^2$ is defined in \eqref{e:J2}. By comparison with the result obtained by means of the correlation function involving the virial current, one comes to the conclusion that $c_T^{\vp} = 1$, in agreement with \eqref{c_coef}. The change in the value of $c_T^{\vp}$ is directly related to a different form of the contribution from the Wess-Zumino action \eqref{e:WZ} as the effect of the reparametrisation of the metric.

\subsection{\texorpdfstring{$\< TTTT \>$}{< TTTT >}}

In case of the 4-point function, the counterpart of the expression \eqref{e:TTTTtowork} reads
\begin{align} \label{e:TTTTtw}
& \< T(\bs{x}_1) T(\bs{x}_2) T(\bs{x}_3) T(\bs{x}_4) \> = \frac{\delta^4 W}{\delta \vp(\bs{x}_1) \delta \vp(\bs{x}_2) \delta \vp(\bs{x}_3) \delta \vp(\bs{x}_4)} \nn\\
& \qquad + \: \left[ \< \frac{\delta^2 S_{\text{int}}}{\delta \vp(\bs{x}_1) \delta \vp(\bs{x}_2)} T(\bs{x}_3) T(\bs{x}_4) \> + 5 \text{ permutations} \right] \nn\\
& \qquad - \: \left[ \< \frac{\delta^2 S_{\text{int}}}{\delta \vp(\bs{x}_1) \delta \vp(\bs{x}_2)} \frac{\delta^2 S_{\text{int}}}{\delta \vp(\bs{x}_3) \delta \vp(\bs{x}_4)} \> + 2 \text{ permutations} \right] \nn\\
& \qquad + \: \left[ \< \frac{\delta^3 S_{\text{int}}}{\delta \vp(\bs{x}_1) \delta \vp(\bs{x}_2) \delta \vp(\bs{x}_3)} T(\bs{x}_4) \> + 3 \text{ permutations} \right].
\end{align}
As we can see the only difference between this expression and \eqref{e:TTTTtowork} - apart from the reparametrisation of the interaction action \eqref{e:Sintvp} - is the change in the sign of the last term. However, the contribution from the Wess-Zumino action reads now
\begin{equation}
\frac{\delta^4}{\delta \vp(\bs{p}_1) \delta \vp(\bs{p}_2) \delta \vp(\bs{p}_3) \delta \vp(\bs{p}_4)} (\delta_\sigma S_{WZ}) = - 12 \sigma e_{TT} \sum_{1 \leq i < j \leq 4} p_i^2 p_j^2.
\end{equation}
This expression vanishes in the on-shell limit as noticed in \cite{Dymarsky}. Therefore, by the optical theorem, one could argue that the 4-point function in \eqref{e:TTTTtw} becomes semi-local in the forward scattering limit. Nevertheless, due to a non-zero value of the constant $c_T^{\vp}$ in the action \eqref{e:Sintvp}, the local terms in \eqref{e:TTTTtw} contribute non-trivially to the scale violation. In particular, with $c_T^{\vp} = 1$ as found in the previous section,
\begin{align}
& \hdelta_\sigma \lla \frac{\delta^2 S_{\text{int}}}{\delta \vp(\bs{p}_1) \delta \vp(\bs{p}_2)} T(\bs{p}_3) T(\bs{p}_4) \rra = - 2 \sigma (c^{\vp}_2)^2 e_{22} (\bs{p}_1 \cdot \bs{p}_2) (\bs{p}_3 \cdot \bs{p}_4) + 2 \sigma e_{TT} \times \nn\\
& \qquad\qquad \times \left[ - (\bs{p}_1 + \bs{p}_2)^4 - p_3^4 - p_4^4 + 2 (\bs{p}_1 + \bs{p}_2)^2 p_3^2 + 2 (\bs{p}_1 + \bs{p}_2)^2 p_4^2 + 2 p_3^2 p_4^2 \right], \\
& \hdelta_\sigma \lla \frac{\delta^2 S_{\text{int}}}{\delta \vp(\bs{p}_1) \delta \vp(\bs{p}_2)} \frac{\delta^2 S_{\text{int}}}{\delta \vp(\bs{p}_3) \delta \vp(\bs{p}_4)} \rra = - 2 \sigma e_{TT} (\bs{p}_1 + \bs{p}_2)^4 - 2 \sigma (c_2^{\vp})^2 e_{22} (\bs{p}_1 \cdot \bs{p}_2) (\bs{p}_3 \cdot \bs{p}_4),\\
& \hdelta_\sigma \lla \frac{\delta^3 S_{\text{int}}}{\delta \vp(\bs{p}_1) \delta \vp(\bs{p}_2) \delta \vp(\bs{p}_3)} T(\bs{p}_4) \rra = - 2 \sigma c_T^{(3)\vp} e_{TT} p_4^4.
\end{align}
When combined, one finds
\begin{align}
& \hdelta_\sigma \lla T(\bs{p}_1) T(\bs{p}_2) T(\bs{p}_3) T(\bs{p}_4) \rra = - 8 \sigma \left( e_{TT} + \tfrac{1}{4} c_2^2 e_{22} \right) \times \nn\\
& \qquad\qquad \times \left[ (\bs{p}_1 \cdot \bs{p}_2) (\bs{p}_3 \cdot \bs{p}_4) + (\bs{p}_1 \cdot \bs{p}_3) (\bs{p}_2 \cdot \bs{p}_4) + (\bs{p}_1 \cdot \bs{p}_4) (\bs{p}_2 \cdot \bs{p}_3) \right] \nn\\
& \qquad - \: 4 \sigma e_{TT} \left( 1 + \tfrac{1}{2} c_T^{(3)\vp} \right) \times \left( p_1^4 + p_2^4 + p_3^4 + p_4^4 \right).
\end{align}
For this expression to agree with \eqref{e:TTTTfromV} one needs $c_T^{(3)\vp} = -2$, in agreement with \eqref{c_coef}.
This calculation confirms the result in the main text.

\section{Multiple scalar operators} \label{sec:multiple}

In this appendix we discuss the case of multiple scalar operators of dimension two and four. Consider a theeory with $n$ scalar operators $\O_2^i$, $i=1,2,\ldots, n$ of dimension two and $N$ scalar operators $\O_4^I$, $I = 1,2,\ldots, N$ of dimension four (in addition to the trace of the stress-energy tensor). All normalisation constants introduced in \eqref{e:TT} - \eqref{e:24} carry now additional indices, \textit{e.g.}, $e_{22}^{ij}$, $e_{4T}^I$ and so on.

First we will argue that by adding an appropriate improvement term one can
make all off-diagonal 2-point functions of the trace of the stress-energy tensor and scalar operators of dimension two and four vanish. To show this let us consider the $(n+N+1)$-dimensional vector space spanned by the independent vectors ordered as follows
\begin{equation} \label{e:base1}
\{ \partial^2 \O_2^1, \partial^2 \O_2^2, \ldots, \partial^2 \O_2^n, \ T, \ \O_4^1, \O_4^2, \ldots, \O_4^N \}.
\end{equation}
Introduce a scalar product given by the matrix of the 2-point functions
\begin{equation}
M = \left( \begin{array}{ccc}
e_{22}^{ij} & e_{2T}^i & e_{24}^{iJ}  \\
e_{2T}^j & e_{TT} & e_{4T}^J \\
e_{24}^{jI} & e_{4T}^I & e_{44}^{IJ} \end{array} \right).
\end{equation}
The matrix $M$ is symmetric and non-negative defined, due to reflection positivity applied to the state
\begin{equation}
| \Psi \> = \alpha^I \O^I_4 (\bs{x}) | 0 \> + \beta^i \partial^2 \O^i_2 (\bs{x}) | 0 \> + \gamma T(\bs{x}) | 0 \>,
\end{equation}
for arbitrary $\alpha^I$, $\beta^i$ and $\gamma$. If $M$ has null vectors, then some of the operators in the basis \eqref{e:base1} are linearly dependent. The operator given by such a linear combination essentially vanishes, due to the reflection positivity condition on its 2-point function. Therefore, we may remove all null operators and assume that $M$ represents a non-degenerate scalar product. Hence, by the Gram-Schmidt orthogonalisation procedure one can find an orthogonal basis
\begin{equation}
\{ \partial^2 \O'^1_2, \partial^2 \O'^2_2, \ldots, \partial^2 \O'^n_2, \ T', \ \O'^1_4, \O'^2_4, \ldots, \O'^N_4 \}
\end{equation}
related to \eqref{e:base1} by the lower-triangular matrix with ones on the diagonal,
\begin{equation} \label{e:matA}
\left( \begin{array}{c} \partial^2 \O'^i_2 \\ T' \\ \O'^I_4 \end{array} \right) =
\left( \begin{array}{ccc} A^i_j & 0 & 0 \\ A_j & 1 & 0 \\ A^i_J & A_J & A^I_J \end{array} \right) \left( \begin{array}{c} \partial^2 \O_2^j \\ T \\ \O_4^J \end{array} \right),
\end{equation}
where $A^i_j$ and $A^I_J$ are lower-triangular square matrices satisfying $A^i_i = A^I_I = 1$ and $A_j$ and $A_J$ are some vectors. Therefore, when the matrix in \eqref{e:matA} is applied to the set of operators $\{ \O_2^i, T, \O_4^I \}$ the resulting operators $\{ \O'^i_2, T', \O'^I_4 \}$ have their off-diagonal 2-point functions vanish by definition, so indeed we could assume
\begin{equation}
e_{22}^{ij} = \delta^{ij} e_{22}^i, \qquad e_{44}^{IJ} = \delta^{IJ} e_{44}^I, \qquad e_{24}^{iJ} = e_{2T}^{i} = e^I_{4T} = 0,
\end{equation}
where $e_{22}^i$ and $e_{44}^I$ are positive constants for any $i=1,\ldots,n$ and $I=1,\ldots,N$.

This construction works in any reflection positive QFT. We now want to show that the orthogonalised operators can be realised by using the following improvement term, generalising \eqref{e:imp},
\begin{equation}
\Delta S = \int \D^4 \bs{x} \sqrt{g} \left[ \sum_i \xi^i \O_2^i \Sigma + \sum_{i,I} \xi'^{iI} \O_2^i \hat{\Box} \phi_0^I + \sum_I \xi''^I \phi_0^I T + \sum_{i,j} \eta^{ij} \phi_2^i \O_2^j + \sum_{I,J} \eta'^{IJ} \phi_0^I \O_4^J \right],
\end{equation}
where $\phi_0^I$ and $\phi_2^i$ denote sources of the original operators $\O_4^I$ and $\O_2^i$. The two additional terms involving the parameters $\eta^{ij}$ and $\eta'^{IJ}$ are responsible for the mixing among the operators of dimension two and four. In this case equations \eqref{e:Timp} - \eqref{e:O4imp} generalise to
\begin{align}
\left( \begin{array}{c} \O_{2 \text{ imp}}^i \\ T_{\text{imp}} \\ \O_{4 \text{ imp}}^I \end{array} \right) =
\left( \begin{array}{ccc} \eta^{ij} + \delta^{ij} & 0 & 0 \\ \xi^i \partial^2 & 1 & 0 \\ \xi'^{iI} \partial^2 & \xi''^I & \eta'^{IJ} + \delta^{IJ} \end{array} \right)
\left( \begin{array}{c} \O_{2}^j \\ T \\ \O_{4}^J \end{array} \right).
\end{align}
By comparing this expression to \eqref{e:matA} we see that we can choose $\xi^i, \xi'^{iI}, \xi''^I, \eta^{ij}$ and $\eta'^{IJ}$
such that we implement the Gram-Schmidt procedure.

The remaining calculations presented in this paper remain valid with obvious changes. The interaction action \eqref{e:Sint} reads
\begin{align}
S_{\text{int}} & = \int \D^4 \bs{x} \left[ \tau T + C_{\mu} V^{\mu} + \phi^i_2 \O^i_2 + \phi^I_0 \O^I_4 + \ldots \right.\nn\\
& \qquad + \: \tfrac{1}{2} \tau^2 \left( c_T T + c'^i_2 \partial^2 \O^i_2 + c^I_4 \O^I_4 \right) + \tfrac{1}{2} c^i_2 (\partial \tau)^2 \O^i_2 \nn\\
& \qquad \left. + \: \tfrac{1}{2} \tilde{c}^i_{2} C_\mu C^\mu \O^i_2 + \ldots \right],
\end{align}
with all $c$ constants acquiring respective indices. Then the following changes follow:
\begin{itemize}
\item In sections \ref{sec:TT2} and \ref{sec:TT4} we discussed 3-point functions involving a single scalar operator. All results remain valid when the appropriate indices are introduced. In particular one finds
\begin{equation}
c'^i_2 = 0, \qquad c_4^I = 0, \qquad c_2^i = \tilde{c}_2^i
\end{equation}
for $i=1,2,\ldots,n$ and $I=1,2,\ldots,N$.
\item The scale violation \eqref{e:TTTT} in the 4-point function of the trace of the stress-energy tensor receives a contribution from all operators of dimension two according to,
\begin{align} \label{e:TTTT}
& \hdelta_\sigma \lla T(\bs{p}_1) T(\bs{p}_2) T(\bs{p}_3) T(\bs{p}_4) \rra = - 8 \sigma \left( e_{TT} + \tfrac{1}{4} \sum_{i=1}^n (c^i_2)^2 e^i_{22} \right) \times \nn\\
& \qquad\qquad \times \left[ (\bs{p}_1 \cdot \bs{p}_2) (\bs{p}_3 \cdot \bs{p}_4) + (\bs{p}_1 \cdot \bs{p}_3) (\bs{p}_2 \cdot \bs{p}_4) + (\bs{p}_1 \cdot \bs{p}_4) (\bs{p}_2 \cdot \bs{p}_3) \right].
\end{align}
Therefore all conclusions we reached in section \ref{sec:TTTT} remain valid.
\end{itemize}

\providecommand{\href}[2]{#2}\begingroup\raggedright\endgroup

\end{document}